\documentclass[fleqn,usenatbib]{mnras}

\usepackage{newtxtext,newtxmath}

\usepackage[T1]{fontenc}
\usepackage{ae,aecompl}

\usepackage{graphicx}	
\usepackage{amsmath}	
\usepackage{amssymb}	

\newcommand{\TFc}{\boldsymbol{\mathcal{F}}} 

\newcommand{\average}[1]{\left \langle {#1} \right \rangle}

\newcommand{\T}{\mathsf{T}} 
\newcommand{\I}{\mathbf{I}} 
\newcommand{\TF}{\text{\bf TF}} 

\newcommand{\CovMat}{\boldsymbol{\Sigma}} 

\newcommand{\D}{\mathbf{G}} 

\newcommand{\dint}{\mathrm{d}} 
 
\newcommand{\thetavec}{{\boldsymbol{\theta}}} 
\newcommand{\alphavec}{{\boldsymbol{\alpha}}} 
 
\newcommand{\rhovec}{\boldsymbol{\rho}}
\newcommand{\kappavec}{\boldsymbol{\kappa}}
\newcommand{\etavec}{{\boldsymbol{\eta}}} 
\newcommand{\phivec}{{\boldsymbol{\psi}}} 
\newcommand{\varphivec}{{\boldsymbol{\varphi}}} 
 
\newcommand{\svec}{{\mathbf{s}}} 
\newcommand{\xvec}{{\mathbf{x}}}

\newcommand{\rvec}{{\mathbf{r}}}

 \newcommand{\FTR}{\widetilde{\mathcal{R}}} 
  \newcommand{\FTG}{\widetilde{\mathcal{W}}} 
  \newcommand{\FTS}{\widetilde{\mathbf{s}}} 
  \newcommand{\FTWF}{\widetilde{\boldsymbol{\varphi}}} 
  \newcommand{\FTN}{\widetilde{\boldsymbol{\eta}}} 

  \newcommand{\Sphi}{\mathbf{W}_\varphi} 

\newcommand{\qed}{\nobreak \ifvmode \relax \else
      \ifdim\lastskip<1.5em \hskip-\lastskip
      \hskip1.5em plus0em minus0.5em \fi \nobreak
      \vrule height0.75em width0.5em depth0.25em\fi}

\title[High-contrast with Pyramid Sensors]{Performance limits of adaptive-optics/high-contrast imagers
with pyramid wave-front sensors}

\author[C. M. Correia et al.]{
Carlos M. Correia,$^{1,2}$\thanks{E-mail: carlos.correia@lam.fr (CMC)}
Olivier Fauvarque, $^{2}$
Charlotte Z. Bond,$^{1}$
\newauthor
Vincent Chambouleyron, $^{2}$
Jean-Fran\c{c}ois Sauvage,$^{2,3}$
Thierry Fusco,$^{2,3}$
\\
$^{1}$WM Keck Observatory, 65-1120 Mamalahoa Hwy Kamuela, HI 96743, USA\\
$^{2}$Aix Marseille Univ, CNRS, CNES, LAM, Marseille, France\\
$^{3}$ONERA, 29 avenue de la division Leclerc, 92322 Chatillon, France
}

\date{Accepted XXX. Received YYY; in original form ZZZ}

\pubyear{2020}

\begin{document}
\label{firstpage}
\pagerange{\pageref{firstpage}--\pageref{lastpage}}
\maketitle

\begin{abstract}
Advanced AO systems will likely utilise Pyramid
wave-front sensors (PWFS) over
the traditional Shack-Hartmann sensor in the quest for increased
sensitivity, peak performance and ultimate contrast. 
Here, we wish to bring knowledge and quantify the PWFS
theoretical limits as a means to highlight its properties and use
cases.
We explore forward models for the PWFS in the
spatial-frequency domain for they prove quite useful since a) they emanate
directly from physical-optics (Fourier) diffraction theory; b)
provide a straightforward path to meaningful error breakdowns, c)
allow for reconstruction algorithms with $O (n\,log(n))$ complexity for large-scale
systems and d) tie in seamlessly with decoupled (distributed) optimal
predictive dynamic control for
performance and contrast optimisation. All these aspects are dealt
with here. 
We focus on recent
analytical PWFS developments and demonstrate the performance using both analytic
and end-to-end simulations.
We anchor our estimates with observed on-sky
contrast on existing systems and then show very good agreement between
analytical and  Monte-Carlo estimates for the PWFS.  For a potential upgrade of existing high-contrast imagers
on 10\,m-class telescopes with visible or near-infrared PWFS, we show under median conditions at Paranal a contrast
improvement (limited by chromatic and scintillation effects) of
2x-5x by replacing the wave-front sensor alone at large separations
close to the AO control radius
where aliasing dominates, and factors in excess of 10x by coupling
distributed control with the PWFS over most of the AO control region,
from small separations starting with the Inner Working Angle of
typically 1-2 $\lambda/D$ to the AO correction edge (here 20 $\lambda/D$).
\end{abstract}

\begin{keywords}
instrumentation: adaptive optics -- methods: analytical -- 
atmospheric effects --
techniques: high angular resolution 
\end{keywords}


\section{The quest for performance and contrast}
The first generation of high-contrast imagers on 10\,m-class
telescopes has been working over the last $~$5 years or so, producing
exquisite images of scattered light from discs in circumstellar environments (\cite{beuzit19, macintosh19, mouillet18, xuan18, guyon18, mawet16}). However, the discovery of new planets
has been quite disappointing with very few confirmed detections.

There are of course no culprits to blame, yet limiting contrast has been raised
as a limitation that shall be lifted in order to populate the
long-waited list of new discoveries (\cite{mawet14, cantalloube19}). In this respect, it has been
recognised that the inner-working angle (IWA) of coronagraphs is to be
decreased to as close as possible to $\lambda/D$ in an attempt to
observe close-in new planets. For such endeavour, novel coronagraph
concepts galore (\cite{guyon18, mawet12, snik18}).

On a par, AO-related residuals can be further reduced in
hopes to improve contrast across the AO correction band (typically up
to a separation of
few tens of $\lambda/D$, depending on the deformable mirror's linear number of actuators) \cite{correia17}. 
As pointed out in \cite{guyon05}, the effects that limit the
performance of wave-front correction are
\begin{enumerate}
\item noise on the WFS (photon - fundamental, read-out -
  technological), requiring sub-electron noise detectors. 
\item Aliasing arising from the discrete nature of the WFS
  measurement, damped with appropriate wave-front sensors
\item Servo-lag error due to the dynamic rejection of residuals in a
  negative feedback loop, calling for faster/more clever
  algorithms
\item actuator fitting, demanding higher-density deformable mirrors
\end{enumerate}
To these adds chromatic optical path length difference (OPD) and amplitude errors
 between the WFS wavelength $\lambda_{WFS}$ and the imaging wavelength
 $\lambda_{im}$  that we revisit and
 fully take into account (\cite{guyon05, fusco06, hardy98}).
 
In this paper we provide AO-limited performance and limiting contrast
estimates 
when a perfect coronagraph is employed. We show the expected improvement
with the use of pyramid WFS in both near-infrared (NIR) and
visible (VIS)
wavelengths with a realistic 2D physical-optics model capable of
mimicking effects as modulation, partial AO correction causing PSF broadening and extended
sources (\cite{fauvarque19}). Additionally we investigate the usefulness of predictive 
control through the application of Kalman filters and distributed
control in the
spatial-frequency domain (\cite{correia17}).

Throughout the paper we use models in the spatial-frequency domain
(Fourier for short) for a number of good reasons, each addressed in a
dedicated section.
\begin{enumerate}
\item physical-optics optical transfer functions are naturally
  described in the Fourier plane -- \S~\ref{sec:opticalModels}
\item statistically independent error terms are readily evaluated from
  the residuals -- \S~\ref{sec:analyticErrorBudgetEval}
\item wave-front reconstruction can be seamlessly done using standard
  filter operations -- giving rise to the use of matrix-free operations
  with $O \left(n\,log(n)\right)$ complexity algorithms for
  large-scale systems - \S~\ref{sec:WFR}
\item allow for decoupled (distributed) optimal filters for
performance and contrast optimisation - \S~\ref{sec:limitPerfContrast}
\end{enumerate}


We assume that non-common path errors are properly corrected for
and thus do not enter the AO-centric error budget developed here.





\section{Optical models of the Pyramid wave-front sensor using diffraction theory}\label{sec:opticalModels}

The behaviour of the Pyramid in the spatial domain has been
extensively studied by~\cite{Verinaud04, verinaud05, chew06,
  korkiakoski07, ledue09, QuirosPacheco09,
  wang10, shatokhina13, fauvarque15, fauvarque17}, following the seminal work of~\cite{ragazzoni96} who builds on the footsteps of Linfoot's Foucault knife-edge
diffraction model~(\cite{linfoot48}).  

Here we stick to the original 4-facet pyramid concept, although
generalisations to any number of facets exist, using coherent or
incoherent recombination of light past the pyramid optic
(\cite{fauvarque15, fauvarque17}). The latter can be designed to optimise contrast at
certain separations, yet the lack of a general design compelled us
with some loss of generality to
consider the original P-WFS concept only.

The intensity pattern on each of the 4 re-imaged pupils at the
detector plane $i_q(\xvec,t)$, $q \in \{1,\cdots,4\}$
indexed by a bi-dimensional coordinate $\xvec = (x,y)$ and time $t$
is conveniently formulated using Fourier masking 
\begin{equation}\label{eq:pyrIntensityFull}
i_q(\xvec',t) = \int_{t-T_s}^t \left| \TFc^{-1}\left\{ H_q(\kappavec)\TFc\left\{
      A(\rvec)
      e^{i\left(\psi(\rvec)+\theta(\rvec,t)\right)}\right\}\star o(\kappavec)\right\}\right|^2
\dint t 
\end{equation}
where $Ae^{i\psi}(\rvec)$ is the electric-field in the pupil
$\mathcal{A}$ (for aperture)
$A(\rvec)$ its amplitude, $\psi(\rvec)$ its phase --
Fraunhofer-propagated to the focal-plane using a 2-D Fourier transform
$\TFc$. This focal-plane field is 2D convolved by the object
$o(\kappavec)$ which has the net effect of a modulation since each
point of the object adds to the phasor $ e^{i\left(\theta(\rvec,t)\right)}$.  $\theta(\rvec,t)$ is an additional time-dependent \textit{modulation} signal,
introduced here as a phase increment to the aberrated wave-front over
the integration time $T_s$ in the pupil-plane $\rvec = (r_x,r_y)$. A customarily used signal is a
time-varying tilt that shifts the focal-plane electric field and makes
it wander across the 4 pyramid facets.  
Next in line, $H_q$ is a masking function (or transparency mask) placed at
the focal plane indexed by $\kappavec
= (\kappa_x, \kappa_y)$ for each $q^{th}$ quadrant of the form
\begin{equation}
H_q (\kappavec) = \mathcal{H}_{\pm \kappa_x}  \mathcal{H}_{\pm \kappa_y}
e^{-i \alphavec_q (\pm \kappa_x) \cdot (\pm \kappa_y)}
\end{equation}
where $\mathcal{H}_{\pm \kappa_x} $ is the Heaviside function for
either positive or negative spatial frequencies and $\alpha_q \in
\mathbb{R}$ a real-valued variable that sets the output angle of
the re-imaged pupils with respect to the chief-ray.
In practice, on a computer, we
replace the integral by a sum on temporally incoherent intensity patterns each
for a tilt value (of the modulation or of the object). The number of sums is calculated based on the
sampling of the PSF at the WFS detector focal plane, although we could
opt to replace this regular sample by irregular sampling, finer across
the pyramid edges and coarser on top of the facets with still
consistent results~(\cite{fauvarque17a}). 

\subsection{Impulse response of a PWFS}

Traditionally the PWFS signals are extracted from the 4 re-imaged pupils
using a slope-like formulation which stems from the original Foucault
knife-edge test. It provides a notional first-derivative measurement of
the wave-front
\begin{equation}\label{eq:SxSy}
\svec_x = g_{opt}^x\frac{i_1 + i_2 - i_3 - i_4}{\sum_q i_q} -s_x^0
\hspace{15pt} \svec_y = g_{opt}^y\frac{i_1 - i_2 + i_3 - i_4}{\sum_q i_q} -s_y^0
\end{equation}
with $g_{opt}^{x,y}$ the optical gain (\cite{bond17a, esposito15})
and $s_{x,y}^0$ the null-phase reference measurement. Henceforth,
this definition is referred to as the \textit{slopes-map} model.

Special care must be paid with respect to the denominator of
\eqref{eq:SxSy}, whether to normalise each value by the sum of the 4
corresponding intensities or by replacing $\sum_q i_q$ by a scalar value
representing the total integrated flux, i.e. $\int_\Omega i_q \dint
\Omega$ with $\Omega$ the domain set by the valid pixels (\cite{Verinaud04}). The latter
is considered a more robust option~(\cite{Bond16}). 

Here we strive to provide a meaningful yet practical, linear physical-optics
forward model of the PWFS that can under minimal simplifications represent
the bulk of its operation, yielding a convolution of the input phase
by the sensor's impulse-response (IR)
\begin{equation}\label{eq:IRslopes}
\svec_x = \mathbf{IR}_{\svec_x}\star \varphivec \hspace{15pt} \svec_y = \mathbf{IR}_{\svec_y}\star \varphivec
\end{equation}

Both \cite{conanr03} and
\cite{shatokhina13} show that the pyramid slopes-map can be
asymptotically
approximated as
\begin{equation}\label{eq:ConvolutionMeasEq}
\svec_x = -\frac{J_0(\alpha x)}{\pi x} \star\Pi_p \star \varphivec(x,y) \hspace{10pt} \svec_y = -\frac{J_0(\alpha y)}{\pi y} \star\Pi_p\star \varphivec(x,y) 
\end{equation}
when the telescope aperture is considered infinite and the phase
aberrations $\varphivec <<1\,rad$. Conan went on to develop
the effect of cross-terms (from adjacent and opposite quadrants of the PWFS), yet we
refrain from using it for the formulation that follows -- \S
\ref{sec:model_vs_measurement} -- proved more
practical, more condensed and therefore less cumbersome. For completion, $J_0(\cdot)$ is a
zero-order Bessel function of the first kind and $alpha$ a real-valued scalar representing the modulation in units of $\lambda/D$;  we have added the function $\Pi_p$ to Conan's
and Shatokhina's to represent the pixel response -- and likewise a
user-defined pixel binning for poor signal-to-noise ratio (SNR)
regimes with dimmer stars --
conveniently modelled as a door function (\cite{oppenheim99}). This term carries the
smearing of the sensitivity curves observed experimentally for spatial
frequencies closer to the system's control radius. Unlike
\cite{Verinaud04} we consider the finite nature of the measurement an
integral part of the sensing chain leading to a different insight into
the nature of the measurements provided by both the PWFS and the
SH-WFS as shown in Fig. \ref{fig:pyramidSensitivitySlices}. The 1D curves are quite insightful for understanding the PWFS behaviour yet rather limited since the 2D sensitivity is far from being radially-symmetric as shown in Fig. \ref{fig:sx_sy_Filters_measured}.

We note that this model, as is happens, can be formulated as a linear
combination of intensity terms each following a more general definition covering cases of the coherent and incoherent recombination of light past the pyramid optic) in
the form
\begin{equation}\label{eq:IRintensity}
i_{linear} = \mathbf{IR}\star \varphivec
\end{equation} 
which admits a
closed-form expression. We follow \cite{fauvarque17} to dub this the
\textit{meta-intensity} model. They show that
\eqref{eq:pyrIntensityFull} can be linearised using a Taylor expansion
series and the Cauchy product of two complex series to circumvent the
squared modulus. This is particularly insightful as they show that recombining the
quadrant intensities as is customary -- see Eq.~(\ref{eq:SxSy}) --
improves the PWFS linearity range 
as the even-powers intensity dependence on the phase cancel out, pushing the non-linearity further
away to higher-order terms. This is so with perfectly aligned systems whereas in practice this assertion may not fully hold (\cite{deo18}) which is a
clear indication to use directly the intensity signals instead of the slopes at the expense of reduced linearity range.

The models provided in \cite{fauvarque17} that we adopt in this study -- as
generalisations which they are --  allow for
different transparency masks with variable number of facets, extended
guide-stars, the effect of the telescope pupil and the presence of residual errors after AO partial
compensation. Moreover, \cite{fauvarque19} develop further the IR
in Eq. (\ref{eq:IRintensity}) reaching an analytic
formulation suitable to the estimation the optical gains from the power-spectral density of
AO residuals.

Equation (\ref{eq:IRintensity}) is very appealing to perform wave-front
reconstruction in that $i_{linear} $ is a closer match to the PWFS
physical-optics model than the model in
Eq. (\ref{eq:ConvolutionMeasEq}) implies.

We chose explicitly to work with improved "slopes-maps'' models for ease of
understanding and comparison to SH-WFS. 

For analytic performance evaluation (in the absence of elements of
practical nature, such as calibration, optical defects, saturation etc) using linear models in the form of either
Eq. (\ref{eq:IRslopes})  or Eq. (\ref{eq:IRintensity}) leads to the
same results. For real-time wave-front reconstruction using directly
detector intensities may lead to computational savings and a more
appropriate setting -- yet we let this discussion open and do not
dwell on it  here.

Next section recasts the formulations seen so far in the  Fourier
domain where the required 
mathematical operations admit simplifications and are soundly and effectively accomplished.

\subsection{Transfer Function of a PWFS}

We now turn our focus into the physical-optics model in the spatial-frequency
domain. This formulation is especially useful
since measurements are obtained as the convolution of the phase by
the PWFS impulse-response, or, equivalently, as a point-wise
multiplication in the Fourier domain. Let the following general-purpose linear measurement model 
  \begin{equation}\label{eq:S_FT}
    \FTS\left(\boldsymbol{\kappa}\right) = \FTG
    \FTWF_{||}\left(\boldsymbol{\kappa}\right) + 
    \widetilde{\boldsymbol{\alpha}}\left(\boldsymbol{\kappa}\right) + 
    \widetilde{\boldsymbol{\eta}}\left(\boldsymbol{\kappa}\right),
\end{equation}
where $\FTG = \left\{\FTG_x; \FTG_y\right\}$ is a linear filter obtained by Fourier transforming the
impulse-response in Eq. \eqref{eq:ConvolutionMeasEq} -- i.e. the
\textbf{PWFS optical transfer-function} (OTF) -- relating the in-band
wave-front  $\FTWF_{||}$ to the 
measurements $\FTS$, $
\widetilde{\boldsymbol{\alpha}}\left(\boldsymbol{\kappa}\right)$ is
the aliasing term acting as a
generalised (coloured) noise term and 
$\widetilde{\boldsymbol{\eta}}\left(\boldsymbol{\kappa}\right) $ is 
additive noise representing photon and detector read noise (\cite{correia14, correia17}).

\begin{figure}
	\begin{center}
            \includegraphics[width=0.5\textwidth]{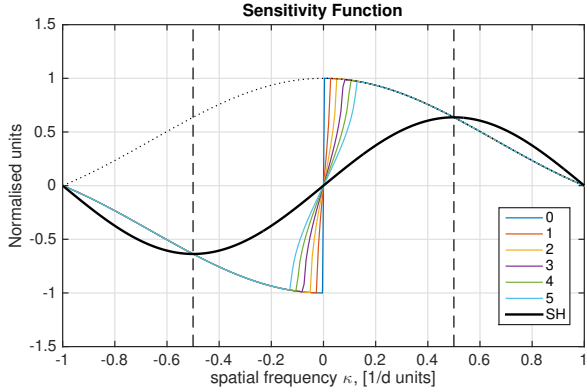}
	\end{center}
	\caption[]
	{\label{fig:pyramidSensitivitySlices}
Pyramid 1D sensitivity plots for modulations $\{0,\cdots,5\}\lambda/D$ from \eqref{eq:pyramidPOmodel} overlaid with
Shack-Hartmann sensitivity when the discrete nature of the measurement
is explicitly taken into account for a fair comparison between the
PWFS and SH-WFS. Although the PWFS exhibits a
slope-like and phase-like measurement regimes, this misconception is clarified in the text.}
\end{figure}

\cite{shatokhina13} provide a closed-form equation
for the Fourier transform of \eqref{eq:ConvolutionMeasEq} which is a
generalisation from the one-dimensional~\cite{Verinaud04}
linear modulation PWFS to the two-dimensional case with
circular modulation, allowing for an expression of the filter as
\begin{equation}\label{eq:pyramidPOmodel}
  \FTG_x(\boldsymbol{\kappa}) = \left\{ 
    \begin{array}{ccc}
      i\, sgn(\kappa_x)sinc(b d\boldsymbol{\kappa}) & \mbox{if $|\boldsymbol{\kappa}|> \kappa_\text{mod}$} \\
      \frac{2i}{\pi}arcsin(\kappa_x/\kappa_\text{mod}) sinc(b d\boldsymbol{\kappa}) 
                                &  \mbox{if $|\boldsymbol{\kappa}|< \kappa_\text{mod}$}
    \end{array} \right.
\end{equation}
with $\boldsymbol{\kappa} = (\kappa_x;\kappa_y)$ a two-dimensional
spatial frequency vector in units of m$^{-1}$, $\kappa_\text{mod}$ the modulation $\alpha$ from Eq. (\ref{eq:ConvolutionMeasEq}) expressed in m$^{-1}$, $d$ the sampling of the pupil plane in meters (commonly the sub-aperture size) and
$\FTG_y(\kappa_x,\kappa_y) = \FTG_x(\kappa_y,\kappa_x) $,
i.e. the transpose of the 'x' filter. Considering that the
notion of discrete averaging at the detector level causes a damping
at high frequencies closer to the AO control radius given by a
multiplicative separable factor $sinc(\boldsymbol{\kappa}) =
sinc(\kappa_x)sinc(\kappa_y)$ with $sinc(x) = sin(x\pi)/(x\pi)$. This
term represents also the user-defined, post-facto binning with $b\in \mathbb{N}$ an
integer scalar. Figure \ref{fig:pyramidSensitivitySlices} depicts 1-D
slices of $\FTG(\boldsymbol{\kappa})$ across the spatial-frequency
variables, representing the sensitivity of the pyramid optic and
integrated with the ensuing (discrete spatial sampler) detector. 

If instead we use developments by \cite{fauvarque19}, then the PWFS OTF
can be formulated as
\begin{equation}\label{eq:ConvolutionMeasEqFauvarque}
\FTG_x(\boldsymbol{\kappa})
=\sqrt{|\widetilde{\mathcal{T}}_x(\boldsymbol{\kappa})  \star \widetilde{\Pi}_p|^2}
\end{equation}
where the function $\widetilde{\mathcal{T}}_x(\boldsymbol{\kappa})$
expands as
\begin{equation}\label{eq:ConvolutionMeasEqFauvarqueExpanded}
\widetilde{\mathcal{T}}_x(\boldsymbol{\kappa})  = 2i \left(H_3\star H_2\omega-H_2\star
  H_3\omega + H_1\star
H_4\omega - H_4\star H_1\omega\right)
\end{equation}
with 
$\omega$ a
weighting function that characterizes the modulation signal,\textit{ i.e.}, it encodes the normalised time spent on the modulation
phase $\thetavec(\rvec,t)$ over one integration frame. Provided it is expanded as a linear series
of $n$ modes, $\thetavec(\rvec,t) = \sum_{i=1}^n a_i(t)\mathcal{M}(\rvec)$
with $\mathcal{M}(\rvec)$ an orthonormal basis set, $\omega$ becomes
\begin{equation}\label{eq:weightSignal}
\omega = \widetilde{\mathcal{A}} \star \int_{t-T_s}^{T_s} a(t)\dint t
\end{equation}
where $\tilde{\mathcal{A}}$ is the Fourier-transformed aperture function.

The most common
choice is tilt modulation for which case we have $\thetavec(\rvec,t) = a_1(t) x + a_2(t)
y$. If the modulation describes a perfect radially-symmetric ring, $a_1 = a_2=$ and consequently, Eq. (\ref{eq:weightSignal}) becomes (\cite{badour11})
\begin{equation}\label{eq:weightSignalSimplified}
\omega = \TF^{-1}\left\{\mathcal{A} (\rhovec) \times J_0(\alpha \rhovec)\right\}
\end{equation}
where $J_0(\alpha \rhovec)$ a Bessel function resulting from the Fourier transform of a (modulation) circle in the focal-plane. The relationship between modulation and tilt amplitude is $\alpha = \pi/4 a$ in units of $\lambda/D$. \textbf{This is the formulation that we will use in the remainder of this paper unless otherwise specified.}

We note from
Eq.~(\ref{eq:ConvolutionMeasEqFauvarque}) that the OTF is in fact the
modulation transfer function (MTF) which provides the magnitude response of the optical system 
to  harmonic functions  of  different  spatial  frequencies. The PWFS phase transfer
function (PTF) is therefore null, a consequence of using intensity
signals to measure a complex field.

A remarkable feature of this model is that the PWFS instantaneous
response can now be understood and potentially used to estimate
instantaneous optical gains (through the use of a complex-valued
$\tilde{\mathcal{A}}$ function to a) optimise the run-time AO
performance and b) estimate (and remove) quasi-static (pinned)
speckles that limit the contrast achievable with high-contrast
imagers.

\subsection{A note on the nature of the PWFS signals} \label{sec:note-nature-pwfs}
The dual behaviour of the modulated 
Pyramid sensor, acting as a \textit{slope-like sensor} for low spatial frequencies and
a \emph{phase-like sensor} for high spatial frequencies is now well
established yet it corresponds to a misconception. It stems from an
erroneous analogy between the sensitivity of the pyramid and the
nature of the measured signal initially stated in \cite{Verinaud04}
and represented in Fig. \ref{fig:pyramidSensitivitySlices}. Although
for spatial-frequencies above the modulation $m\times\lambda/D$ the sensitivity is that of a
phase sensor, the PWFS provides still a signal akin to
the first spatial
derivative of the phase (in the form of a Hilbert transform) with a frequency-dependent scaling at the
origin of the misconception. Dubbing the PWFS measurements as
"slopes'' is, under this light, justified. Paradoxically, it is commonplace in
the AO community. 

In theory, for a large modulation the sensor will act more fully as a gradient
sensor (with the correct frequency-dependent gains) and it may be
possible to reconstruct from its measurements using previously derived
Shack-Hartmann filters by \cite{correia14}. One such successful albeit
sub-optimal attempt can be found in~\cite{QuirosPacheco09}.

\subsection{Modelled vs. measured PWFS filter functions} \label{sec:model_vs_measurement}
\begin{figure}
	\begin{center}
          \includegraphics[width=0.25\textwidth]{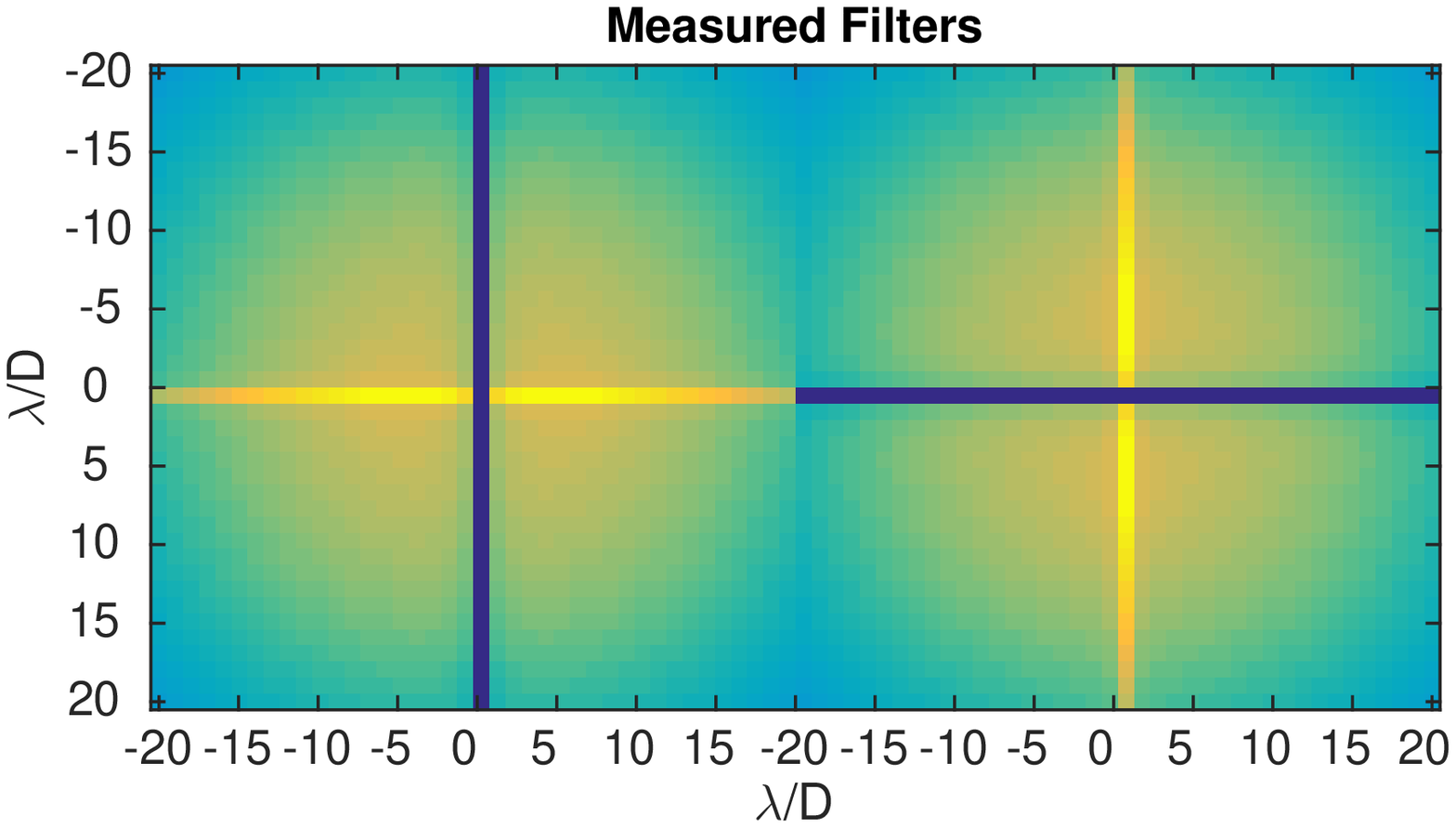}\includegraphics[width=0.25\textwidth]{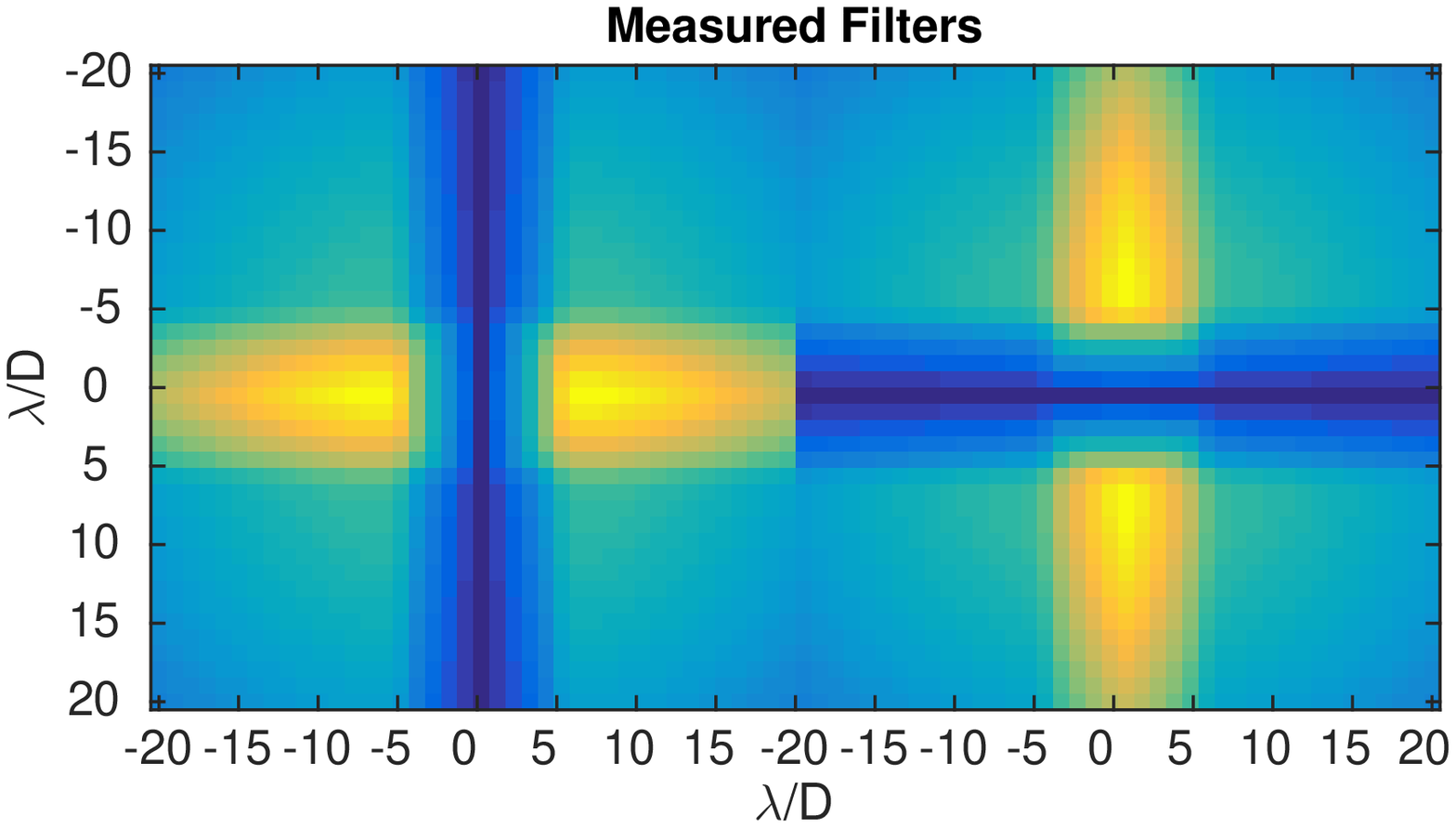}
                    \includegraphics[width=0.25\textwidth, height=0.12\textwidth]{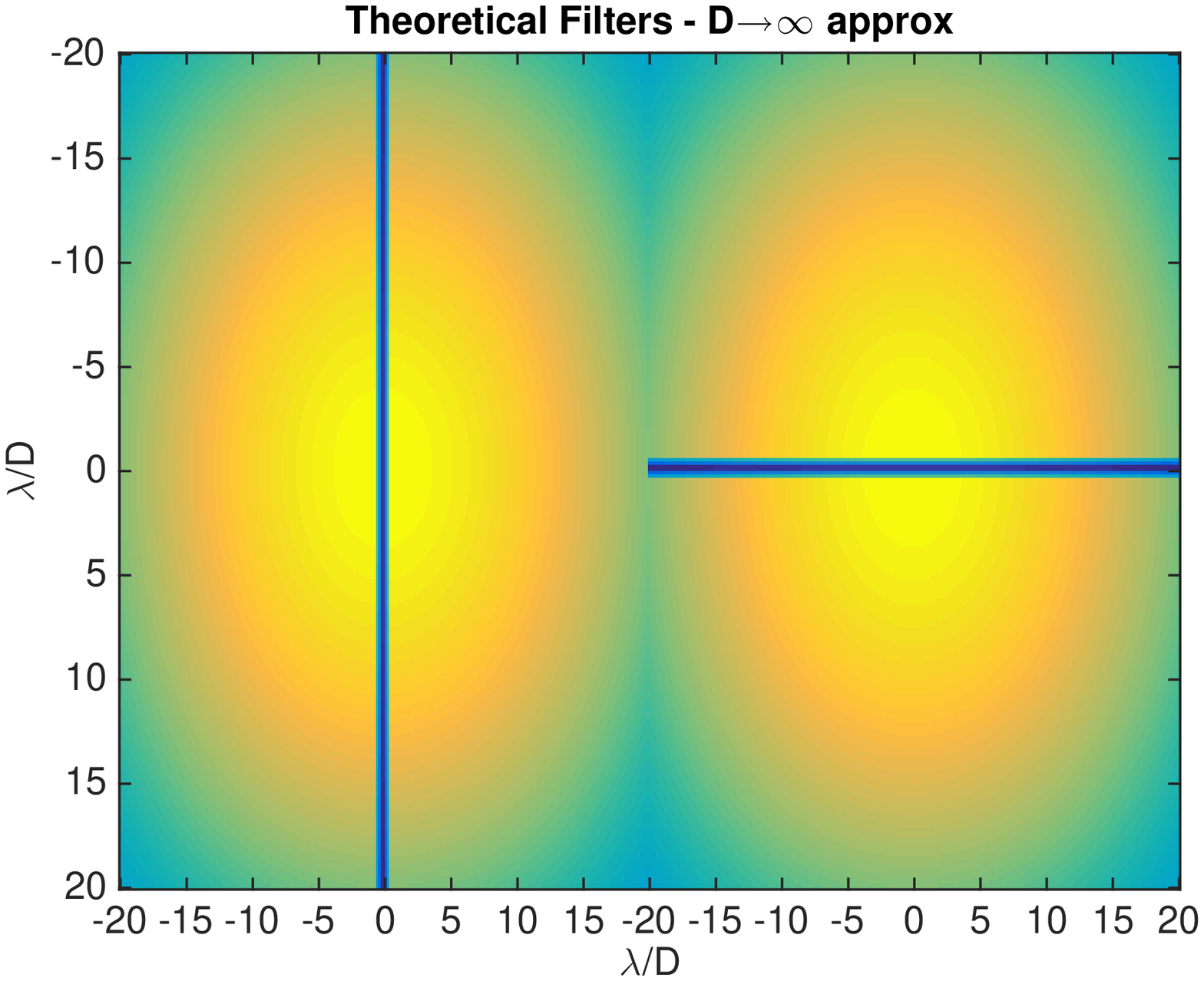}\includegraphics[width=0.25\textwidth, height=0.12\textwidth]{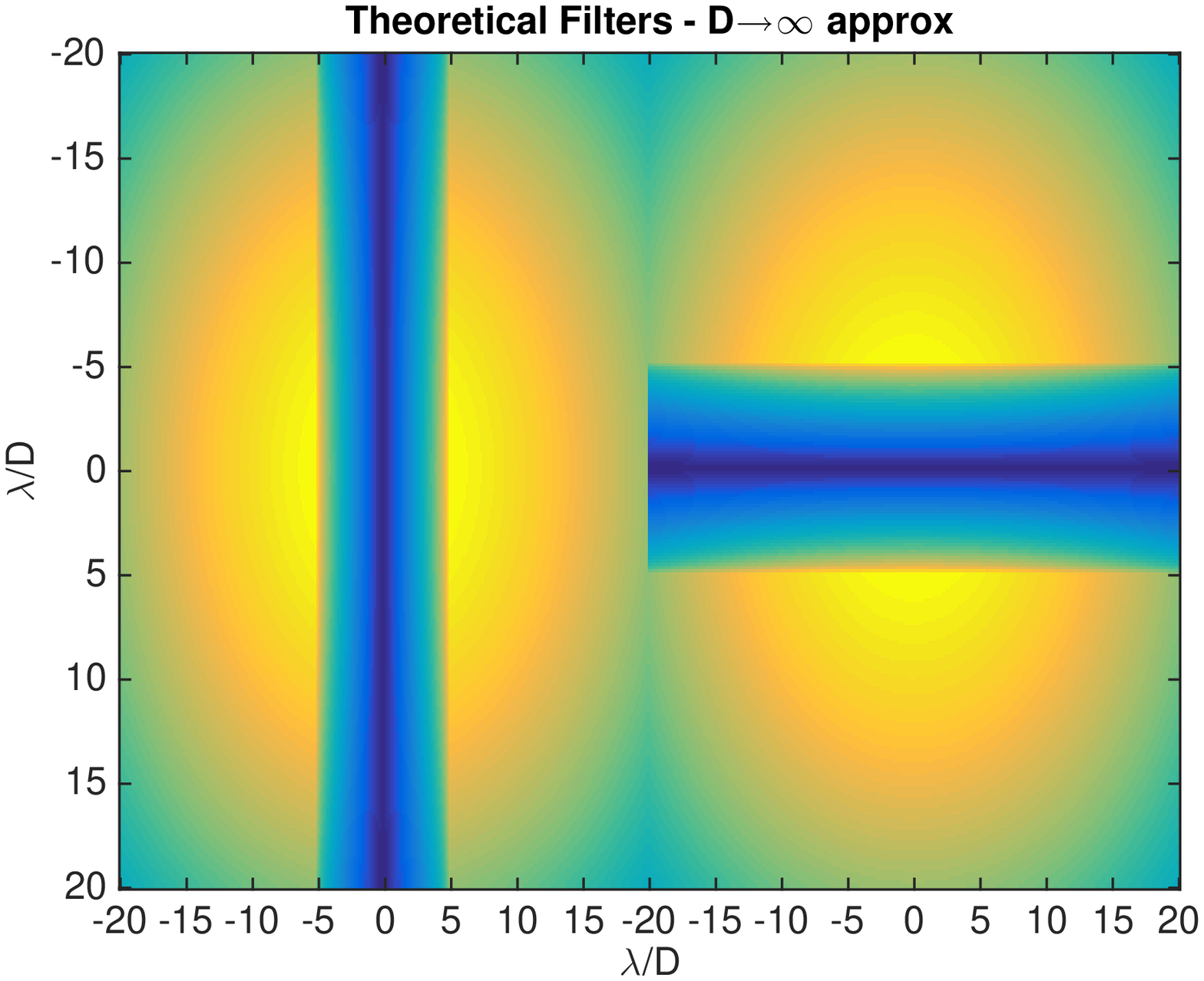}
            \includegraphics[width=0.25\textwidth]{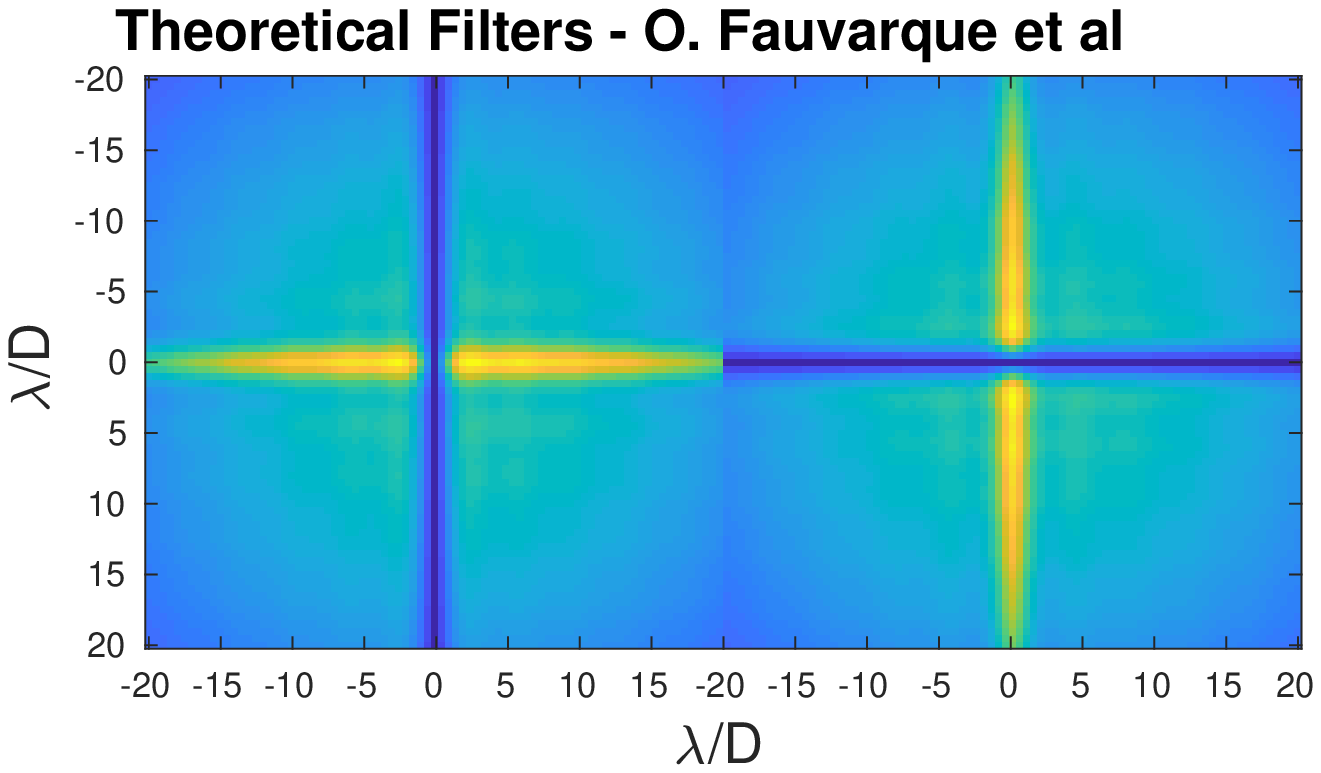}\includegraphics[width=0.25\textwidth]{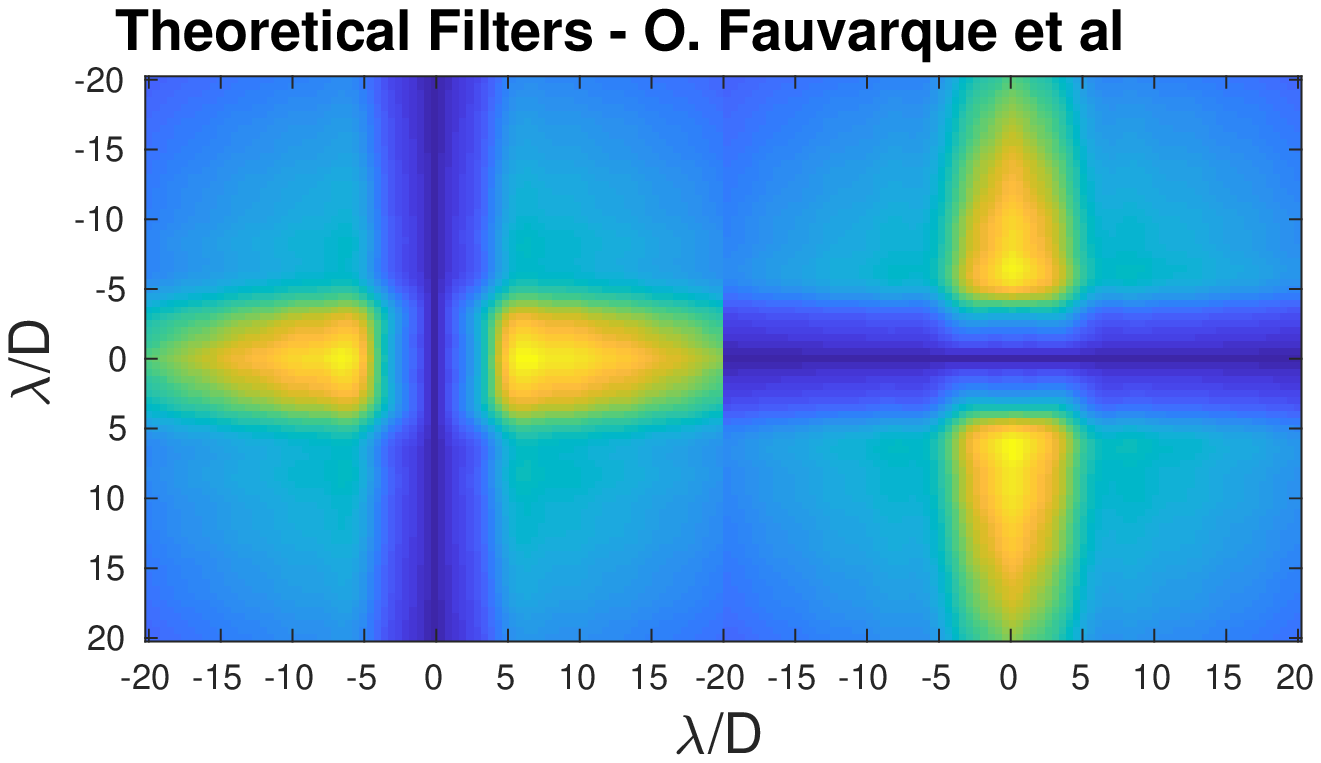}
	\end{center}
	\caption[]
	{\label{fig:sx_sy_Filters_measured}
Comparison of measured with theoretical filters. Top:
OOMAO-provided measurement (implementing
\eqref{eq:pyrIntensityFull}) . Mid: model from
Eq. \eqref{eq:pyramidPOmodel}. Bottom: Model from Eq. \ref{eq:ConvolutionMeasEqFauvarque}. Left: modulation $1\lambda/D$. Right:
$5\lambda/D$. }
\end{figure}

Figure \ref{fig:sx_sy_Filters_measured} shows the measured PWFS OTF
using a full end-to-end physical optics model in \cite{conanr14} \textit{OOMAO}
 implementing Eq. \eqref{eq:pyrIntensityFull}. The procedure is reminiscent of the "poke-matrix" in that we record the P-WF response to the complete set of complex-exponential functions in our basis set. It is
compared to the model in Eq. (\ref{eq:pyramidPOmodel}) (which does not
take into account the cross-terms for ease of presentation, although formulated in \cite{conanr03, wang10}) and in
Eq. \eqref{eq:ConvolutionMeasEqFauvarque} for which one can clearly see the
correct fit to the low-, high- and cross-term frequencies.


\subsection{PWFS measurement noise model}
In \cite{feeney01}  it is established that the effect of photon noise (yet not limited to)
on the WFS measurements is such that
\begin{equation}\label{eq:sigmaMeasurementPWFS}
\sigma^2_{\svec_x, photon} = \sum_q \sigma^2_{i_q} \left(
  \frac{\partial \svec_x}{\partial i_q}\right)^2
\end{equation}
where $\sigma^2_{i_q}$ is the signal variance on the $q^{\text{th}}$ quadrant and
\begin{equation}
\svec_x = f(i_1,i_2,i_3,i_4)
\end{equation}
from Eq. (\ref{eq:SxSy}). The PWFS diffracted field gives rise to local intensity variations in
the re-imaged pupil planes leading to
$\sigma^2_{i_q} = \average{i_q}$ under Poisson statistics where
$\average{\cdot}$ stands for ensemble-averaging. Besides, light falls outside the
valid re-imaged pupils most prominently for low modulation cases, leading to a loss of SNR. Although one such
noise model taking into account these features can be obtained
straightforwardly, it lacks simplicity and practicality. We will assume
for the sake of simplicity that the number of incident photons on each pixel is the same
yielding
\begin{equation}\label{eq:sigmaImagePWFS}
\sigma^2_{i_q} \approx n_{ph}/4
\end{equation}
where $n_{ph}$ is the average number of photon detections on the PWFS.

Assuming the raw measurement
\begin{equation}
\rvec_x = \frac{i_q + i_n}{i_q + i_d}
\end{equation}
with $i_d$ and $i_n$ the shorthand for the other-than-$q$ quadrant
intensities in the denominator and numerator respectively of Eq. (\ref{eq:SxSy}).
The  measurement partial derivatives are readily found
\begin{equation}\label{eq:dsdi}
\frac{\partial \svec_x}{\partial i_q} = \frac{\partial \rvec_x}{\partial i_q} \frac{\partial \svec_x}{\partial \rvec_x} = \frac{i_q + i_d + i_q + i_n}{(i_q+i_d)^2}\frac{\partial \svec_x}{\partial \rvec_x}
\end{equation}
with $\frac{\partial \svec_x}{\partial \rvec_x}$ playing the role of
the pixel-dependent optical gain $g_{opt}$. For brevity and practical
reasons, we assume it to be a scalar value.
Using Eq. (\ref{eq:sigmaImagePWFS})
\begin{equation}\label{eq:drdi}
\frac{\partial \rvec_x}{\partial i_q} = \frac{4n_{ph}/4}{n_{ph}^2} = \frac{1}{n_{ph}}
\end{equation}

Plugging \eqref{eq:drdi} into \eqref{eq:dsdi} and then into \eqref{eq:sigmaMeasurementPWFS}, one finally gets
\begin{equation}\label{eq:sigmaPhNoiseMeasurementPWFS_final}
\sigma^2_{\svec_x, photon} = \left[4
  \frac{n_{ph}}{4}\left(\frac{1}{n_{ph}}\right)^2 \left(\frac{\partial \svec_x}{\partial \rvec_x} \right)^2\right] = \frac{g_{opt}^2}{n_{ph}}
\end{equation}

For the read-out noise, following the same assumptions, 
\begin{equation}
\sigma^2_{\svec_x, ron} = \sum_q \left(
  \frac{\partial \svec_x}{\partial i_q}\right)^2
\end{equation}
yielding
\begin{equation}\label{eq:sigmaRONMeasurementPWFS_final}
\sigma^2_{\svec_x, ron} = 4g_{opt}^2\left(\frac{ron}{n_{ph}}\right)^2
\end{equation}
where $ron$ is the average read-out-noise in photo-electrons per frame and per pixel.

\begin{figure}
	\begin{center}
            \includegraphics[width=0.5\textwidth]{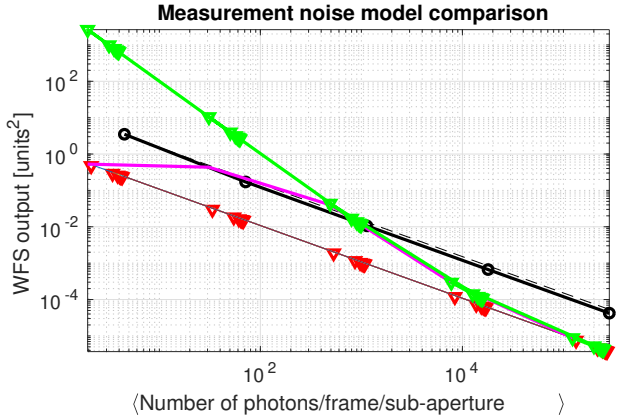}
	\end{center}
	\caption[]
	{\label{fig:photonAndRead50e-Noise_nL16s}
Illustrative example of noise models for the PWFS. Markers: theoretical from expressions in this section, no-markers:
from physical-optics simulation models embedded in \textit{OOMAO} (\cite{conanr14}). Red: photon-noise only. Green: photon+read noise. Magenta: physical-optics for photon+read noise. The saturation at low photon-count is the non-linear measurement regime manifesting (intensities are always positive). SH-WFS curves in black. 
} 
\end{figure}

Figure \ref{fig:photonAndRead50e-Noise_nL16s} compares the different models to physical-optics simulations, showing the great accuracy of the noise (photon and read) models (photon  -
$\sigma^2_\eta$ - and read-noise
(50e-) - $\sigma^2_{ron}$. The multiple markers correspond to different modulations since the photo-electron count varies accordingly, albeit slightly.

\textbf{Comparison to the SH-WFS:} Since the PWFS has often been (wrongly)
likened to the SH in quad-cell mode we provide the general expressions
for the latter to enable the comparison offered in
Fig. (\ref{fig:photonAndRead50e-Noise_nL16s}).

From \cite{thomas06}
\begin{equation}
\sigma^2 = A\frac{1}{\bar{n}_{ph}} + B
\left(\frac{ron}{\bar{n}_{ph}}\right)^2 
\end{equation}
with $A$ and $B$ a function of the algorithm used. $\bar{n}_{ph}$ is the
number of photo-electrons/sub-aperture/frame and $ron$ is the effective read-out noise
in photo-electrons rms. 


For a quad-cell
\begin{align}
A & = \pi^2\kappa\\
B & = 4\pi^2\kappa^2
\end{align}
with $\kappa=1$ for a diffraction-limited spot. 

We observe that using the quad-cell noise model from the SH-WFS
applied to the PWFS leads to results different from those developed
in Eq. (\ref{eq:sigmaPhNoiseMeasurementPWFS_final})  and Eq. (\ref{eq:sigmaRONMeasurementPWFS_final}).

We show in Appendix the necessary steps to calculate the noise
propagation expressed on an orthogonal basis of modes. Although for
the SH-WFS the output variance can be assimilated to an OPD at the edges of
the sub-apertures, we caution that the same cannot be achieved with
the P-WFS on account of the nature of its measurement -- see \S \pageref{sec:note-nature-pwfs}.

\subsection{Considerations about VIS v. NIR WF Sensing}

\textbf{Photometric argument:} Let the accuracy of the measurement be proportional to the diffraction
$\lambda/D$ -- i.e. take the optical gain in
Eq. (\ref{eq:dsdi}) to be inversely proportional to
the angular size of PSF (its full-width at half maximum) that would be recorded at the vertex of the PWFS. A back of the envelope calculation tells us that it is more
beneficial to use NIR wave-front sensing should the number of photons
$n_{ph,NIR} > (\lambda_{NIR}/\lambda_{VIS})^2 n_{ph,VIS}$ 
This standing, a factor $~4$
more photons
is required in the NIR than in the VIS. Taking the case of SPHERE~\cite{fusco16},
the photometric budget for VIS and NIR detectors
(CCD220 and Saphira respectively) coupled with the throughput of the whole
instrument, we get $2\times10^9 ph/s/m^2$ at the central
$\lambda_{VIS}=800\,nm$ whereas at $\lambda_{NIR}=1300\,nm$ we get
$8\times10^9\,ph/s/m^2$ with a G5 star. This seems to indicate that
there is no huge gain in performing NIR WFSensing, avoiding further
operational overheads of operating in the IR.

\noindent\textbf{Morphological argument:} As far as the PWFS is
concerned, since we are not measuring the
position of a PSF (as is the case of the SH-WFS), the previous argument is
flawed, at least to the extent that the PWFS in normal operating
regime features a mixed slope-like and phase-like sensitivities -- \S \ref{sec:note-nature-pwfs} . Inasmuch as the relationship of the morphology of the PSF and
its optical gain is non-linear, operating in the NIR is advantageous
for the residuals at $\lambda_{NIR}$ are lesser allowing the PWFS
  to work closer to its linear regime. Conversely, in the VIS, the
  electric field is way more distorted (greater wave-front residuals),
  causing the PWFS to work in a "less linear'' regime, therefore
  originating optical gain variations that are not fully compensated
  by an increase in photon collection from those sources (\cite{bond18}).

\section{Analytical error budget evaluation} \label{sec:analyticErrorBudgetEval}
\subsection{AO-induced OPD effects}


Results in this section follow closely those in
\cite{correia17}. There we made a comprehensive presentation of
how calculations of aniso-servo-lag, aliasing, measurement noise and
fitting error can be conveniently evaluated using power-spectral
densities in the spatial-frequency domain under temporally-filtered,
closed-loop control. 

\textbf{Throughout this document the parameters in Table
\ref{tab:simBaselineParms} are used by default.}

Our goal is to evaluate the residual (piston-removed) phase
 variance, defined by
  \begin{equation}\label{eq:reconstruction_error}
    \sigma^2_\text{Tot} \triangleq \int
     \widetilde{\mathcal{P}} \left\langle \left| \FTWF(\boldsymbol{\kappa}) - {\FTWF}^{\text{cor}}(\boldsymbol{\kappa})  \right|^2 \right\rangle
    \partial \boldsymbol{\kappa}
  \end{equation}
which is a function of $\{d,D,r_0,L_0,\sigma^2_\eta \}$,
the
actuator pitch, the telescope diameter, the atmosphere coherence
length, the outer scale and the measurement noise variance.
The piston-removal function is given by $\widetilde{\mathcal{P}} =
\left[ 1-\left| \frac{2 J_1(\pi \kappavec D)}{\pi \kappavec
      D}\right|^2\right]$with the term within the module the Fourier
transform of a circular pupil function of diameter $D$.

In the remainder we suppose that the DM corrects entirely for the
reconstructed phase, \textit{i.e.}
${\FTWF}^\text{cor}(\boldsymbol{\kappa}) =
\widehat{\FTWF}(\boldsymbol{\kappa}) = \FTR\FTS(\boldsymbol{\kappa})$ when the anti-folding filter is
applied \cite{correia14}

Equation (\ref{eq:reconstruction_error}) is expanded using \eqref{eq:S_FT} yielding
  \begin{equation}\label{eq:errorBudget}
\begin{split}
     \widetilde{\mathcal{P}}\average{\left| \FTWF(\boldsymbol{\kappa})
          - \widehat{\FTWF}(\boldsymbol{\kappa})  \right|^2}
   &  =
   \widetilde{\mathcal{P}}\average{\left|\tilde\varepsilon_\perp(\boldsymbol{\kappa})
     \right|^2} + \widetilde{\mathcal{P}}\average{\left|\tilde\varepsilon_{||}(\boldsymbol{\kappa})
     \right|^2}  \\ & 
= \average{|\FTWF_\perp|^2} \\ &\quad
    + \left|1 - \FTR \FTG\right|^2 \widetilde{\mathcal{P}} \average{\FTWF(\boldsymbol{\kappa})\FTWF(\boldsymbol{\kappa})^* }
    \\&\quad+ \mathbf{W}_\text{RA} \\ & \quad
    + \average{ \widetilde{\mathcal{P}} \left|\FTR \FTN\right|^2} 
\end{split}
  \end{equation}
  with $
  \average{|\FTWF_\perp|^2}$ the PSD of the fitting error (where we approximate
  $ \widetilde{\mathcal{P}}(\boldsymbol{\kappa}) = 1$ for $|\boldsymbol{\kappa}|>1/(2d)$).
  The term 
\begin{equation}
\left|1 - \FTR \FTG\right|^2 \widetilde{\mathcal{P}} \average{\FTWF(\boldsymbol{\kappa})\FTWF(\boldsymbol{\kappa})^* }=  \left|1 - \FTR \FTG\right|^2\Sphi'(\boldsymbol{\kappa})
\end{equation}
 is  the PSD of the open-loop phase reconstruction error and 
  \begin{equation}
    \mathbf{W}_\text{RA}
= \widetilde{\mathcal{P}} \sum_{\mathbf{m}\neq 0}  \left|\FTR(\boldsymbol{\kappa}) \FTG(\boldsymbol{\kappa}+ \mathbf{m}/d)\right|^2 \Sphi(\boldsymbol{\kappa}  + \mathbf{m}/d)
  \end{equation}
  is the PSD of the reconstructed aliasing error.  Finally  
\begin{equation}
\mathbf{W}_\eta = \average{ \widetilde{\mathcal{P}} \left|\FTR \FTN\right|^2}
\end{equation} 
is the PSD of the propagated noise. This model can be (and was) further
generalised to the closed-loop regime by \cite{correia17} when factoring in spatio-temporal
functions characteristic of the loop filtering into
Eq. (\eqref{eq:errorBudget}).

\begin{figure}
	\begin{center}
        \begin{tabular}{c}
            \includegraphics[width=0.35\textwidth]{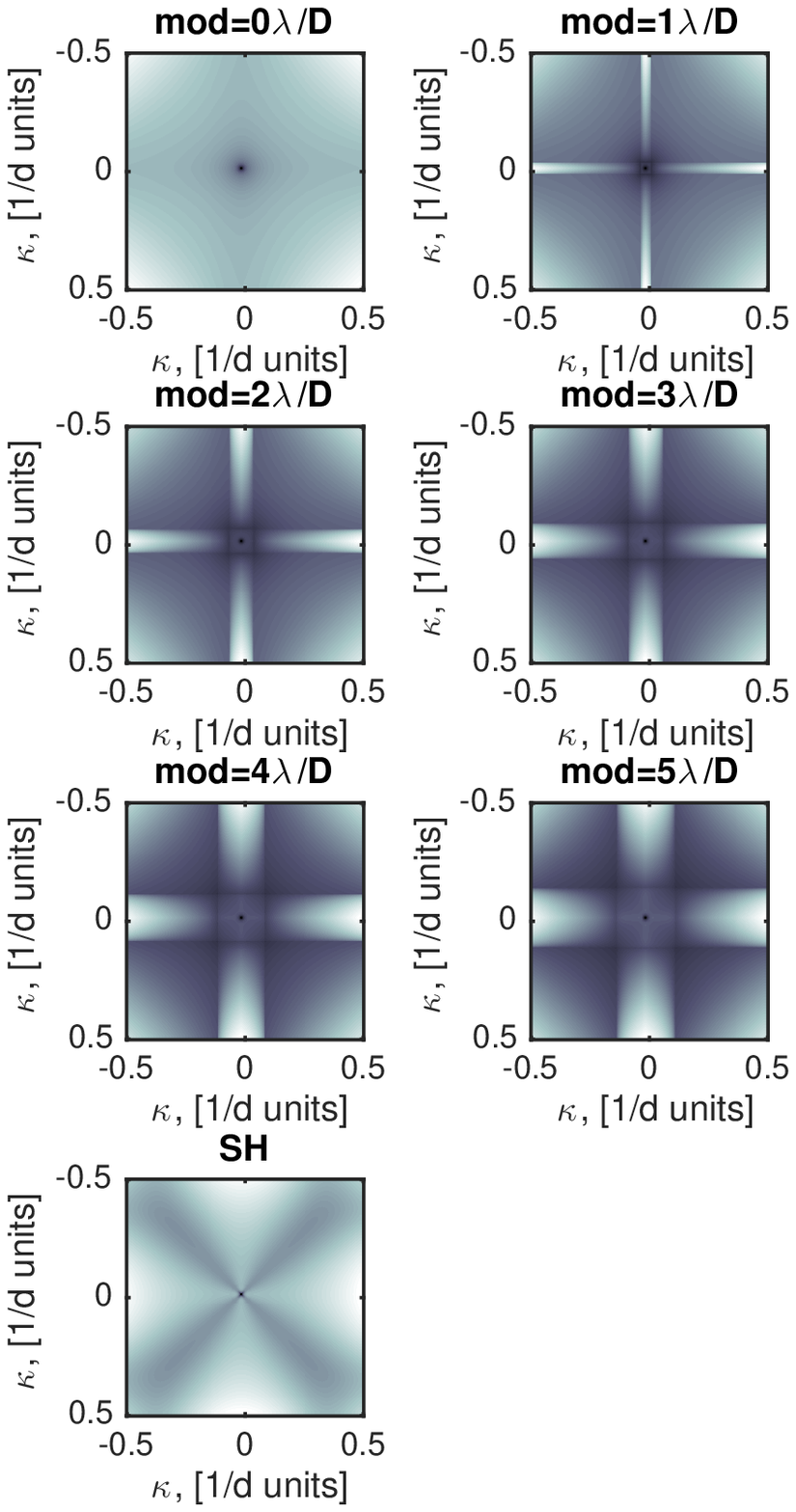}
            \\
            \includegraphics[width=0.4\textwidth]{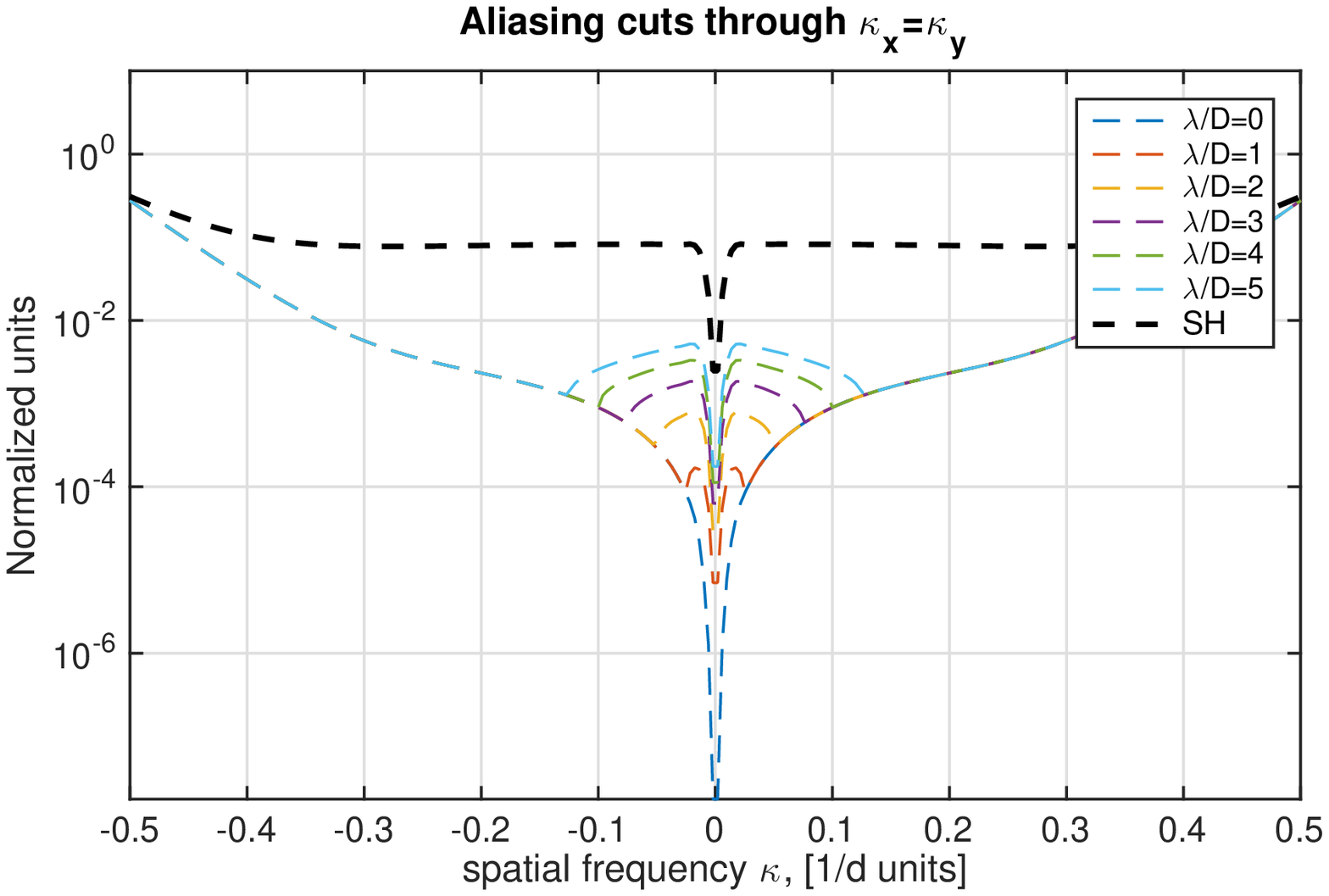}
        \end{tabular}
	\end{center}
	\caption[]
	{\label{fig:aliasKy0Cuts}
Propagated aliasing through LS reconstructor. Top: face-on patterns. Bottom: Slabs $\kappa_y=\kappa_x$. 
}
\end{figure}

\noindent\textbf{Aliasing rejection:} Figure \ref{fig:aliasKy0Cuts}
shows the
propagated aliasing after least-squares wave-front reconstruction
(\cite{correia17}) (\textit{i.e.} with no
temporal loop filtering). When comparing it to Fig. 9 in \cite{Verinaud04}, we
note the general agreement. However, due to the two-dimensional
reconstruction and the way x- and y- frequencies are mixed in the
reconstructor's denominator, a slab along $\kappa_y=0$ or $\kappa_x=0$
shows a damping for very low modulations. A cross-check on the likelihood of
the result is also shown from the cuts along $\kappa_x= \kappa_y$ where all the
propagated aliasing terms for both the pyramid WFS and the SH-WFS
reach the same value at the edge of the control radius.

In either case
the SH-WFS term is provided in black curves for comparison. Its
amplitude is always greater than the one of the PWFS.
The face-on patterns provide further insight into the propagated
aliasing and its spatial distribution. 

From Fig. (\ref{fig:pyramidSensitivitySlices}) one hints at the fact that the amount of aliasing affecting either the PWFS or the SH is about the same. However, it is the propagation through the reconstructor that proves more beneficial with the PWFS.

\begin{figure}
	\begin{center}
            \includegraphics[width=0.5\textwidth]{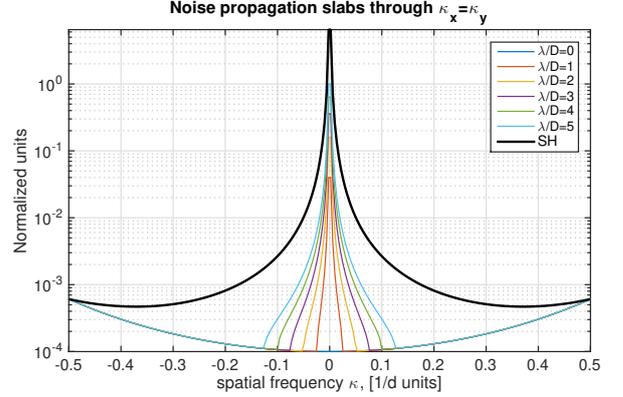}
	\end{center}
	\caption[]
	{\label{fig:noiseDiagCuts}
Propagated noise through LS reconstructor. Slab $\kappa_y=\kappa_x$.}
\end{figure}

\noindent\textbf{Noise propagation:} The noise is propagated through a LS
reconstructor as shown in Fig. \ref{fig:noiseDiagCuts} (compare to
\cite{Verinaud04}, Fig. 7). Here we are just showing a diagonal slab
through the 2D noise propagation filter but the same comment applies
to $\kappa_x=0$ of $\kappa_y=0$ slabs. As for the aliasing, the noise
propagation is always lower for the pyramid WFS across all the spatial
frequencies within the control radius. 

\subsection{Chromatic effects and scintillation}
In addition to AO-induced residual OPD effects, now we consider additional limits to contrast due to chromatic
effects and scintillation which generate both amplitude and OPD
variations. 
\cite{guyon05} and \cite{fusco06} provide
quantification of such effects. We revisit those calculations and provide what we hope a more
comprehensive taxonomy.

A wavelength-dependent index of refraction gives rise to 3 different
errors (\cite{hardy98}), to state,
\begin{enumerate}
\item angular dispersion due to the angle of incidence
and refraction as the field propagates through the medium which can be
seen as a cumulative version of Snell's law over the vertical path
\item  differential refraction error: chromatic path-length and amplitude errors for different wavelengths
traversing the same path for they travel at different velocities creating a
chromatic effect (dubbed correction chromatism
by \cite{fusco06}, term C6 in \cite{guyon05})
\item dispersion displacement error caused by
  differential bending of wave-fronts at different wavelengths, causing rays to probe slightly different patches of turbulence resulting in an angular anisoplanatism-like error
  \end{enumerate}

To these adds scintillation plus OPD due to Fresnel
propagation, even in the absence of chromatic refraction
(\cite{guyon05}) and despite the weak-turbulence regime (\cite{roddier81}).

The first effect listed -- angular dispersion -- has no impact on contrast
but solely on the angular positions of point
sources on the focal plane; it is therefore disregarded in what ensues.

Now, let the index of refraction fluctuations for standard pressure
and temperature from \cite{edlen66} (later slightly adjusted by \cite{owens67})
\begin{equation}
  \Delta n(\lambda) = 8.34213\times 10^{-5}+\frac{0.0240603}{130-\lambda^{-2}} + \frac{0.00015997}{38.9-\lambda^{-2}}
\end{equation}
here taken to coincide with the refractivity index, i.e. $n = 1+\Delta n$.

We assume the differential refraction error to be proportional to the
ratio of fluctuations  $ \Delta
n(\lambda_1)/  \Delta n(\lambda_0)$ as suggested by \cite{fusco06} and not to the ratio of indices of
refraction $n(\lambda_1)/
n(\lambda_0)$ as was considered by
\cite{guyon05}.

For the dispersion displacement error we follow
\cite{fusco06} to compute an error PSD (both OPD and amplitude)
considering anisoplanatic imaging with an angle 
\begin{equation}
    \theta =\Delta
n(\lambda_1) - \Delta n(\lambda_0) tan(ZA)
\end{equation}
where ZA is the zenith
angle in radians.

\begin{figure}
	\begin{center}
         \includegraphics[width=0.53\textwidth]{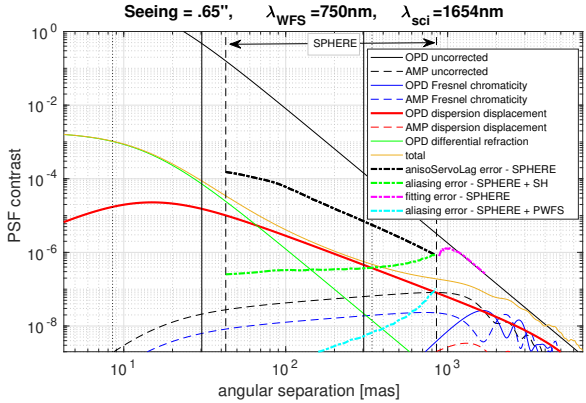} 
	\end{center}
	\caption[]
	{\label{fig:psfContrast_1DCurve_WfsI2_SciH}
Post-coronagraphic PSF contrast curves including atmospheric dispersion effects and
Fresnel propagation on both amplitude (AMP) and phase (optical path length - OPD) for a WFS in the VIS and a camera in the H-band. 
The maximum improvement is comprised between the black dashed curve (the anisoServoLag error) and the orange curve indicating the total of the non-AO limiting contrast terms. In the range \{200\,mas--1000\,mas\} the gap has an upper bound of a factor 10x.
Vertical black lines indicate the correction
band of Keck's AO system (solid), SPHERE (dash) and the ELT (dotted).}
\end{figure}

\begin{figure}
	\begin{center}
         \includegraphics[width=0.53\textwidth]{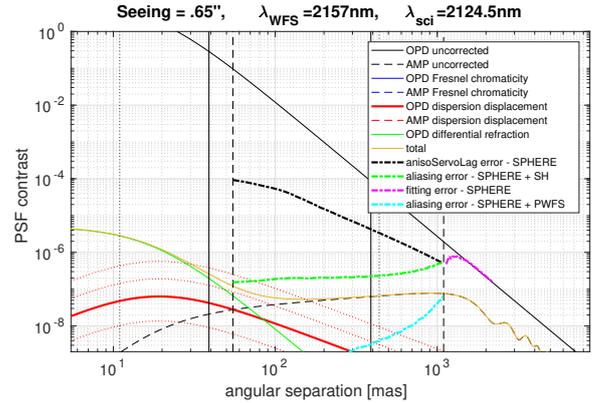} 
	\end{center}
	\caption[]
	{\label{fig:psfContrast_1DCurve_WfsKs_SciKp}
Post-coronagraphic PSF contrast curves including atmospheric dispersion effects and
Fresnel propagation on both amplitude (AMP) and phase (optical path length - OPD) for both K-band WFSensing and imaging. 
The maximum improvement is comprised between the black dashed curve (the anisoServoLag error) and the orange curve indicating the total of the non-AO limiting contrast terms. In the range \{50\,mas--1000\,mas\} the gap reaches two orders of magnitude.
Vertical black lines: see Fig. \ref{fig:psfContrast_1DCurve_WfsI2_SciH} caption.}
\end{figure}

Using standard expressions for Fresnel propagation (real and imaginary
components) and differential chromaticity from \cite{guyon05} we
produced Fig. \ref{fig:psfContrast_1DCurve_WfsI2_SciH} with visible WFS and
NIR imaging and Fig. \ref{fig:psfContrast_1DCurve_WfsKs_SciKp} with
both NIR WFS and imaging. As expected, amplitude effects are more pronounced when using much different wave-front
sensing and imaging wavelengths.

In our implementation we do not consider interference between
servo-lag OPD errors and scintillation although they give rise to tangible effects in
post-coronagraphic images, in particular asymmetric halos in the AO correction region (commonly
known as the butterfly due to its shape) caused by a combination of
temporal delay and wind velocity (\cite{cantalloube18}).

We further notice (as done elsewhere by \cite{guyon05,fusco06,guyon18}) that the AO residuals -- namely the servo-lag error -
is by far the limiting factor. This provides compelling motivation
for the investigation of predictive control approaches -- see for
instance \cite{correia17, males18, massioni15}. We note
also that the contrast estimates are well in-line with the results
obtained with SPHERE on the VLT (\cite{cantalloube19, vigan19})
and Keck (\cite{xuan18}).


\section{Wave-front reconstruction in the spatial-frequency
  domain}\label{sec:WFR}

Having developed PWFS formulations and evaluated the AO-centric error budget from functions in the continuous spatial-frequency
domain in previous sections, we devote this section to the real-time wave-front
reconstruction from pyramid signals using discrete
    deconvolution-based processing as
    a natural extension of the preceding results. The default filter used here is the one recorded by measuring the PWFS response to the Fourier basis set with its physical-optics, diffractive model from Eq. (\ref{eq:pyrIntensityFull}).

The use of Fourier reconstruction has the potential to significantly increase the
reconstruction speed (or otherwise lessen the computational burden), particularly for high-order systems such
as those on Giant Segmented Mirror Telescopes (GSMTs) (\cite{Poyneer05, correia07}).  On the other hand, and
admitting that reconstruction speed in no longer of first priority due
to the huge progress in real-time architectures over the last decade,
we note that the use of spatial-frequencies extends the AO-correctable
area by a factor $\pi/4$ with respect to orthonormal modes defined on
a circular pupil that do not correct frequencies beyond $\kappa =
\sqrt{\kappa_x^2 + \kappa_y^2} > \frac{1}{2d}$ (although one such orthonormal basis could be formulated, to the authors knowledge it has not been used in the past). This ratio can even go
higher when one considers the IWA of the coronagraph that is both
affected by residual diffraction effects and by poor performance of
the coronagraph. 

Extensive development of Fourier reconstruction methods has focused on the
Shack-Hartmann sensor, with compelling results - the most prominent
being the Gemini Planet Imager Fourier Domain Reconstructor (\cite{Poyneer05}).

Advanced systems such as upgrades to existing telescopes (e.g. VLT's
SPHERE, Gemini's GPI, Subaru's ScExAO and Keck's KPIC)
and AO systems for future Giant Segmented Mirror Telescopes are likely to utilise the
Pyramid wave-front sensor over the more commonly used Shack-Hartmann.

A discrete version of the analytic model provided in
Eq. (\ref{eq:pyramidPOmodel}) can be applied to the real-time wave-front
reconstruction from pyramid slope data. An initial
attempt was made in \cite{QuirosPacheco09} assuming a PWFS
sensitivity function to be that of a SH-WFS (which of course is only
valid in the highly modulated case). An implementation customized to
the PWFS is
presented in \cite{shatokhina17} using 1-D reconstruction from PWFS
signals in x and y directions and then averaging. 

Here instead we follow on
the footsteps of \cite{bond17a} and on Fourier-domain implementations
in \cite{correia07, Correia08} that use jointly the x and y measurement data and
take special care of the finite aperture edge effects and boundary conditions using circularity and
divergence of the gradient field to ensure compatibility with the
Fourier series. We find that this treatment is key to obtaining high
levels of performance with minimal losses compared to the case that
uses the full physical-optics PWFS model and SVD filtering to compute
the reconstructor.

\subsection{Discrete pyramid filter filters}

We make use
of a minimum-mean square minimisation criterion to find the following filters 
\begin{equation}
  \widehat{\tilde\phivec} = \mathbf R^x[k,l] \tilde\svec_x +  \mathbf R^y[k,l] \tilde\svec_y
\end{equation}

\begin{equation}\label{eq:Hudgin_enhanced_filter}
  \widehat{\phivec}[k,l] = \left\{ \begin{array}{ccc}
      0 & \mbox{if $k,l = 0$}; \\
      \hat{\mathbf R^x}[k,l] \tilde{\svec}_x[k,l] +\hat{\mathbf R^y}[k,l]\tilde{\svec}_y[k,l]
      & \mbox{otherwise}
    \end{array} \right.
\end{equation}
It is straightforward to demonstrate that the MMSE (Wiener Filter)
writes (\cite{correia14})
\begin{equation}\label{eq:Phi_map}
  \widehat{\mathbf\phivec} = \frac{\mathbf Q_x^* \tilde {\svec}_x
    + \mathbf Q_y^* \tilde{\svec}_y}{|\mathbf Q_x|^2
    + |\mathbf Q_y|^2 + \gamma \frac{\mathbf W_n}{\mathbf W_\phivec}}.
\end{equation}
where $\mathbf Q$ is the discrete-version of the continuous model in
Eq. (\ref{eq:pyramidPOmodel})
with the filters of the form
\begin{equation}\label{eq:Filter_x}
  \hat{\mathbf R^x}[k,l] = \frac{\mathbf Q_x^*}{|\mathbf Q_x|^2
    + |\mathbf Q_y|^2 + \gamma \frac{\mathbf W_n}{\mathbf W_\phivec}} 
\end{equation}
\begin{equation}\label{eq:Filter_y}
  \hat{\mathbf R^y}[k,l] = \frac{\mathbf Q_y^*}{|\mathbf Q_x|^2
    + |\mathbf Q_y|^2+ \gamma \frac{\mathbf W_n}{\mathbf W_\phivec}} 
\end{equation}
The least-squares solution is easily obtained by taking $\gamma=0$.

The priors
$\mathbf W_n$ and $\mathbf W_\phivec = 0.49 r_0^{-5/3}\left\{(2\pi)^2\left[f_x^2 +
    f_y^2 + (1/L_0)^2\right]\right\}^{-11/6}$, are
the spatial PSDs of the noise and the phase. The noise is
assumed white and uncorrelated, thus constant over all the
frequencies, i. e. $\mathbf W_n\propto k \in \Re$. An anti-aliasing
Wiener filtering solution is developed in \cite{correia14a} by
suitably modifying the whiteness of the noise. A further scalar
factor $\gamma$ is introduced to properly
weigh the priors term to account for other unknown system parameters. 

\section{Limiting performance and contrast}\label{sec:limitPerfContrast}

We are now in a position to apply the results from the preceding
sections to representative cases of high-contrast imagers. In this section
we investigate (analytically \& with Monte-Carlo models) the performance for our simulated
system as a function of exposure time, modulation and guide-star
magnitude 
In so doing we
revisit the work of \cite{Verinaud04} and extend it to the
2-dimensional case.  

Further
parameters can be found on Table \ref{tab:simBaselineParms}.
\begin{table}
\caption
{Default simulation parameters. The turbulence model represents median Paranal conditions. Bold represents the nominal conditions.}
\vskip 2mm
\begin{center}
\begin{tabular}{ll}
\hline \hline
{\bf Telescope} & \\
D & 8.0\,m \\
throughput & 50\%\\
\hline
\bf{Guide-star} & \\
zenith angle & \textbf{0}-60 deg \\
magnitude & 0-12\\
\hline
{\bf Atmosphere} & \\
$r_0$ &  15\,cm \\
$L_0$ & 25 m \\
Fractional $r_0$ & [53.28;1.45;3.5;9.57;10.83; \\ & 4.37;6.58;3.71;6.71]/100\\
Altitudes & [0.042;0.140;0.281;0.562;1.125;\\&2.25;4.5;9;18]\,km\\
wind speeds & [15;13;13;9;9;\\&15;25;40;21]\,m/s \\
wind direction & [38;34;54;42;57;\\&48;-102;-83;-77]*$\pi$/180\,deg \\
\hline
{\bf Wave-front Sensor} & \\
Order & 40$\times$40 \\
RON & 1\,e$^-$ \\
$n_{pix}$ &  4\\
$f_{sample} = 1/T_s$ & 0.1--\textbf{1}--5\,kHz \\
modulation $m$ & 0--\textbf{2}--6 $\lambda/D$\\
$\lambda_{WFS}$ & \textbf{0.64}--1.65--2.2\,$\mu$m \\
Centroiding algorithm & thresholded CoG\\
\hline
{\bf DM} & \\
Order & 41$\times$41 \\
\hline
{\bf AO loop} & \\
pure delay  & $\tau_\mathsf{lag}=$1\,ms \\
loop gain  & $g=\{0.01, \cdots, 0.5\}$ \\
\hline
{\bf Imaging Wavelength} & \\
$\lambda_{im}$ & 0.75--\textbf{1.65}--2.2\,$\mu$m \\
\hline
\end{tabular}
\end{center}
\label{tab:simBaselineParms}
\end{table}

\subsection{$\lambda_{im}$=NIR, $\lambda_{WFS}$=VIS in imaging mode} 
\begin{figure}
	\begin{center}
            \includegraphics[width=0.5\textwidth]{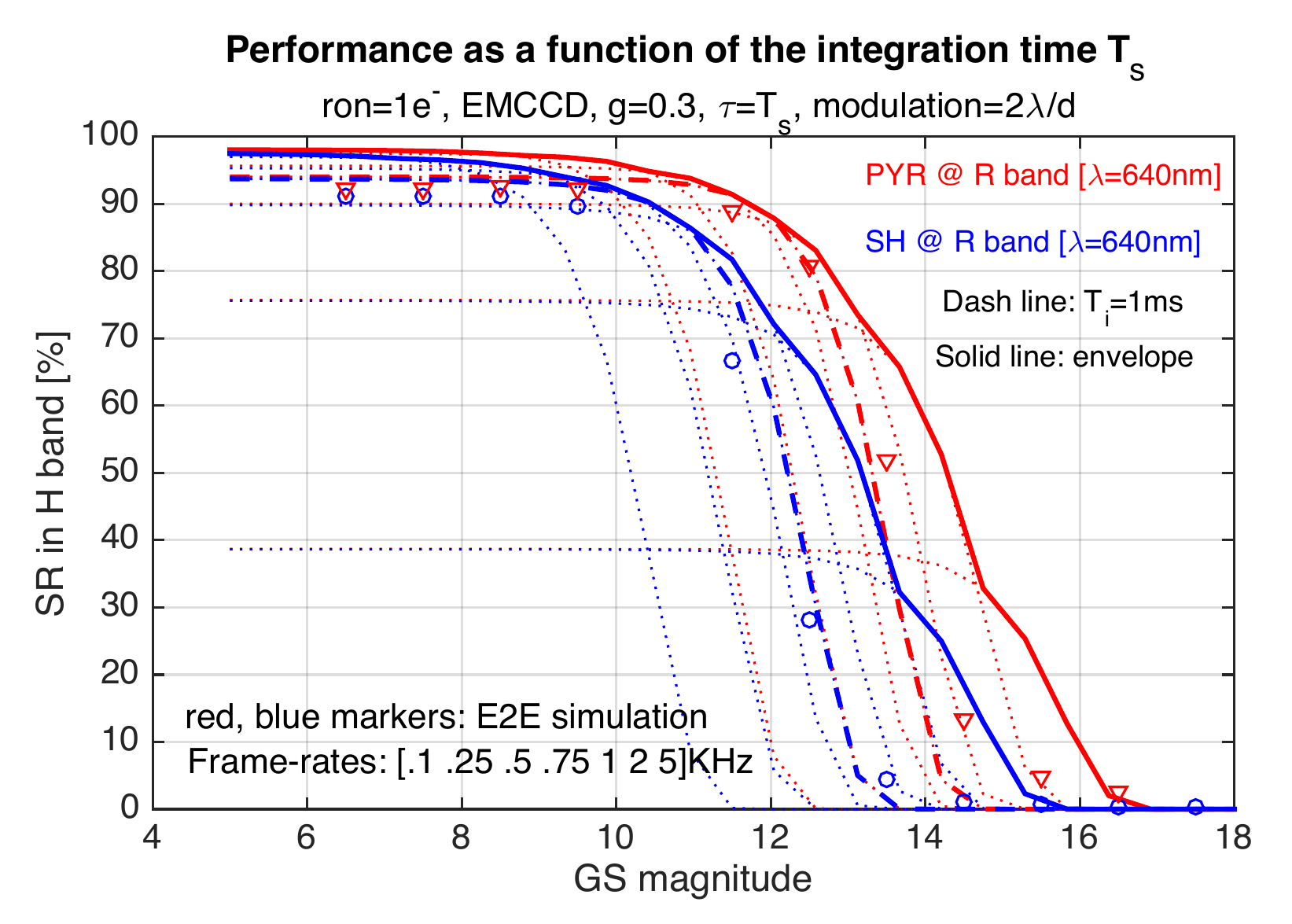}
	\end{center}
	\caption[]
	{\label{fig:strehlVsMag}
Performance comparison between the analytic and the Monte-Carlo
models, both available from \text{OOMAO}. Dashed curves indicate the 1kHz
frame-rate case to be compared to the results (markers only) of the
Monte-Carlo simulations. In the latter the PWFS optical gain are
compensated for by adjusting a single scalar gain which in our opinion
is at the origin of the slight drop in performance on the bright star end.}
\end{figure}

As an example of the analytic error breakdown and reconstruction accuracy,
Fig. \ref{fig:strehlVsMag} compares the limiting performance expected
from a visible PWFS and SH-WFS on a 
8\,m-class telescope as a function of guide-star magnitude 
and AO loop sampling frequency. Using developments in Sect. \ref{sec:WFR} we over-plot (circle and triangle markers) the results of Monte-Carlo simulations performed under \cite{conanr14}'s \textit{OOMAO}.  

It is interesting to note that for bright
guide-stars there is only a slight advantage for the pyramid. 
Looking at the full 2-D
PSFs would further give insight into differences between these two
wave-front sensors that the SR is incapable of showing. 

On the faint star end, the pyramid is at its best. The lower noise
propagation, \textit{i.e.} increased sensitivity, makes it push the
limiting magnitude by roughly two stellar magnitudes. Whereas the SH
drop-off knee is around magnitude 12, the pyramid ensures good
performances down to magnitude 14. At magnitude 15, the pyramid still
achieves $30\%$ SR, a pretty high value.

\subsection{$\lambda_{im}$=VIS, $\lambda_{WFS}$=NIR in imaging mode}





We can likewise explore the performance at visible wavelengths of a
pyramid-based high-contrast AO system. The motivation is two-fold: i)
provide performance in a parameter space complementary to that of
space-borne and ground-based high-resolution spectrographs used for
the indirect detection and characterisation of extra-solar planets and disks and ii) guide
on IR stars and brown-dwarfs, the latter relatively fainter at visible
wavelengths yet likely to host planetary systems as well. 

\begin{figure}
	\begin{center}
            \includegraphics[width=0.53\textwidth]{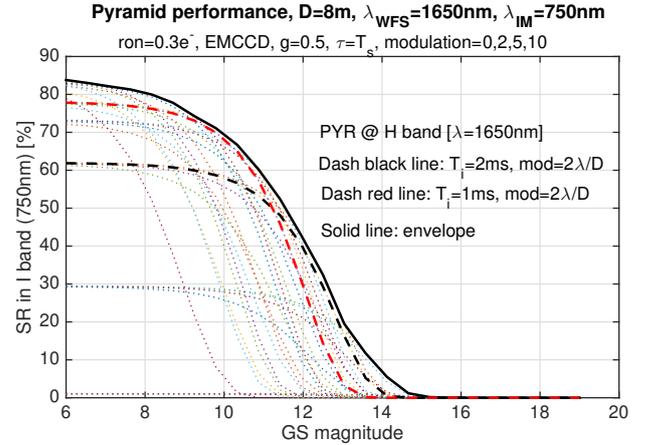}
        \end{center}
	\caption[]
	{\label{fig:sphere2p0ExpectedPerf}
Performance expected at visible wavelengths by a PWFS system.}
\end{figure}

Figure \ref{fig:sphere2p0ExpectedPerf} shows the performance in the I-band ($\lambda_{im}$=850\,nm) when the sensing is done in H-band ($\lambda_{WFS}$=1650\,nm) as a function of
guide-star magnitude, frame-rate and pyramid modulation. We have
considered fast IR detectors with sub-electron noise providing for
fast reading. We can observe the huge impact of running at higher
frame-rates, with the performance increasing from 60\% at 500\,Hz to
80+\% at 5000\,Hz. This is strong indication that servo-lag error is
the dominant factor.

\subsection{Integral vs. distributed control in coronagraphic mode}

In an attempt to minimise residual AO errors after correction, the
results that follow build on the predictive capabilities of
distributed Kalman filters with the formulation presented in~\cite{correia17}.

\begin{figure}
	\begin{center}
            \includegraphics[width=0.55\textwidth]{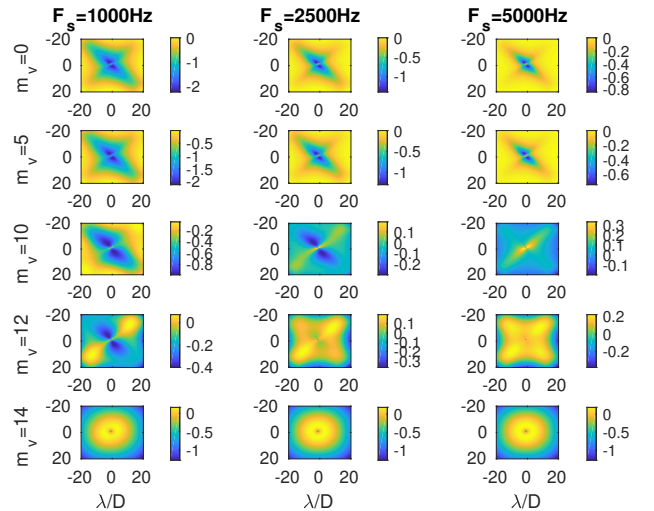}
	\end{center}
	\caption[]
	{\label{fig:sphere2p0ContrastRatio_SH_DKF_vs_SH_LS}
Contrast ratio map (log scale) as a function of the frame-rate and stellar
magnitude for a SH with a LS reconstructor and integral controller
versus a DKF controller.}
\end{figure}

We first investigate the net effect of using predictive control over
single-gain integral control in
Fig. \ref{fig:sphere2p0ContrastRatio_SH_DKF_vs_SH_LS} for a SH-based high-contrast imager. There, we plot the contrast improvement as a ratio (negative values, since we are using a log scale) or a degradation (positive values). 
The 2D nature
of these plots adds to the limiting contrast curves in
Fig. \ref{fig:psfContrast_1DCurve_WfsI2_SciH} and
\ref{fig:psfContrast_1DCurve_WfsKs_SciKp} which radially symmetric in nature;
we can observe on the bright star end that contrast improvements can
surpass the factor 10x at small separations (typically below 5
$\lambda/D$) and in certain wind-dependent directions. 

\begin{figure}
	\begin{center}
            \includegraphics[width=0.55\textwidth]{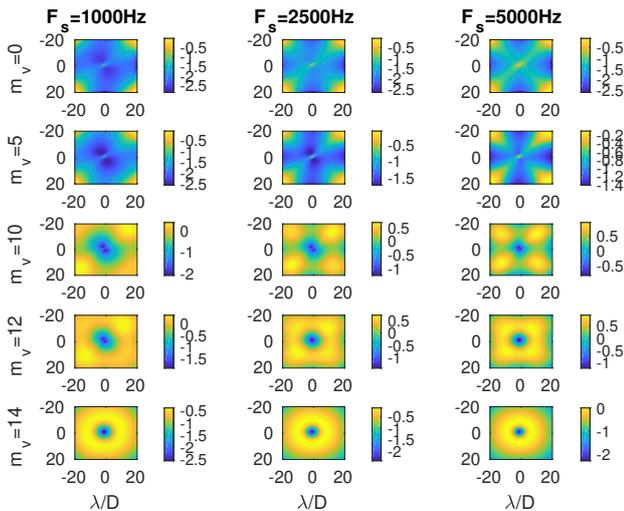}

	\end{center}
	\caption[]
	{\label{fig:sphere2p0ContrastRatio_PWFS_DKF_vs_SH_LsInt}
Contrast ratio map (log scale) as a function of the frame-rate and stellar
magnitude for a SH with a LS reconstructor and integral controller
versus a PWFS with a DKF controller.}
\end{figure}

When a PWFS is employed, then the potential contrast improvement
increases as shown in 
Fig. \ref{fig:sphere2p0ContrastRatio_PWFS_DKF_vs_SH_LsInt} in both depth and extent.

For a magnitude 0 star, a combination of pyramid+DKF can clean up the
AO control region almost entirely, resulting in an
extended zone with up to
two orders of magnitude contrast ratio improvement. This happens on
account of the almost complete removal of servo-lag (typically
$\in \pm 5\lambda/D$) and aliasing terms (closer to the AO correction edges). As we move towards higher
frame-rates, the improvement brought upon by the DKF is lesser for the
servo-lag error is smaller (as in Fig. \ref{fig:sphere2p0ContrastRatio_SH_DKF_vs_SH_LS}).

In real observations the presence of quasi-static speckles can limit
the contrast figures provided. With the possibility offered by the
models in \S \ref{sec:opticalModels} to estimate instantaneously the
PWFS optical gains, we left for the interested reader the exploration
of pushing the contrast further when using noiseless detectors (or
very low read noise) from which high-quality short-exposure images can
be collected with custom post-processing techniques ingrained with
knowledge of the variability of those optical gains.

\section{Conclusion}
We have shown the performance limits of the pyramid wave-front sensor for both imaging and high-contrast applications. For that we produced an AO-centric error breakdown using a practical, convolution-based PWFS model in the spatial-frequency domain (developed in \cite{fauvarque19}) featuring some highly desirable properties
\begin{itemize}
\item its meta-intensity linear model (from which the slopes-maps
  are computed as linear combinations) represents to a broader extent the diffractive nature the PWFS optic. 
  \item this model can be generalised to finite pupils, coherent and incoherent recombination of light resulting from overlapping and non-overlapping re-imaged pupils respectively, 
  extended-objects, and off-line optical gains retrieval
\end{itemize}
Our calculations back the generally accepted result whereby the PWFS extends by up to two stellar magnitudes the limiting WFS magnitude 
  allowing for a larger sky-coverage (not quantified here)
  
On existing high-contrast imagers mounted
on 10\,m-class telescopes with visible or near-infrared PWFS under median Paranal turbulence conditions outlined in \ref{tab:simBaselineParms}, we show a contrast
improvement (limited by chromatic and scintillation effects) of
2x-10x by replacing the wave-front sensor alone at large separations
close to the AO control radius
where aliasing dominates, and factors in excess of 10x by coupling
distributed control with the PWFS over most of the AO control region,
from small separations starting with the Inner Working Angle of
typically 1-2 $\lambda/D$ to the AO correction edge.




\begin{thebibliography}{99}
\makeatletter
\relax
\def\mn@urlcharsother{\let\do\@makeother \do\$\do\&\do\#\do\^\do\_\do\%\do\~}
\def\mn@doi{\begingroup\mn@urlcharsother \@ifnextchar [ {\mn@doi@}
  {\mn@doi@[]}}
\def\mn@doi@[#1]#2{\def\@tempa{#1}\ifx\@tempa\@empty \href
  {http://dx.doi.org/#2} {doi:#2}\else \href {http://dx.doi.org/#2} {#1}\fi
  \endgroup}
\def\mn@eprint#1#2{\mn@eprint@#1:#2::\@nil}
\def\mn@eprint@arXiv#1{\href {http://arxiv.org/abs/#1} {{\tt arXiv:#1}}}
\def\mn@eprint@dblp#1{\href {http://dblp.uni-trier.de/rec/bibtex/#1.xml}
  {dblp:#1}}
\def\mn@eprint@#1:#2:#3:#4\@nil{\def\@tempa {#1}\def\@tempb {#2}\def\@tempc
  {#3}\ifx \@tempc \@empty \let \@tempc \@tempb \let \@tempb \@tempa \fi \ifx
  \@tempb \@empty \def\@tempb {arXiv}\fi \@ifundefined
  {mn@eprint@\@tempb}{\@tempb:\@tempc}{\expandafter \expandafter \csname
  mn@eprint@\@tempb\endcsname \expandafter{\@tempc}}}

\bibitem[\protect\citeauthoryear{Baddour}{Baddour}{2011}]{badour11}
Baddour N.,  2011, \mn@doi [Advances in Imaging and Electron Physics - ADV IMAG
  ELECTRON PHYS] {10.1016/B978-0-12-385861-0.00001-4}, 165, 1

\bibitem[\protect\citeauthoryear{{Beuzit} et~al.,}{{Beuzit}
  et~al.}{2019}]{beuzit19}
{Beuzit} J.~L.,  et~al., 2019, \mn@doi [\aap] {10.1051/0004-6361/201935251},
  \href {https://ui.adsabs.harvard.edu/abs/2019A&A...631A.155B} {631, A155}

\bibitem[\protect\citeauthoryear{Bond, El~Hadi, Sauvage, Correia, Fauvarque,
  Rabaud, Neichel  \& Fusco}{Bond et~al.}{2016}]{Bond16}
Bond C.~Z.,  El~Hadi K.,  Sauvage J.~F.,  Correia C.,  Fauvarque O.,  Rabaud
  D.,  Neichel B.,   Fusco T.,  2016, in AO4ELT-IIII.

\bibitem[\protect\citeauthoryear{Bond, Correia, Sauvage, Neichel  \&
  Fusco}{Bond et~al.}{2017}]{bond17a}
Bond C.~Z.,  Correia C.~M.,  Sauvage J.-F.,  Neichel B.,   Fusco T.,  2017,
  \mn@doi [Opt. Express] {10.1364/OE.25.011452}, 25, 11452

\bibitem[\protect\citeauthoryear{{Bond} et~al.,}{{Bond} et~al.}{2018}]{bond18}
{Bond} C.~Z.,  et~al., 2018, in \procspie. p. 107031Z,
  \mn@doi{10.1117/12.2314121}

\bibitem[\protect\citeauthoryear{{Cantalloube} et~al.,}{{Cantalloube}
  et~al.}{2018}]{cantalloube18}
{Cantalloube} F.,  et~al., 2018, \mn@doi [Astronomy \& Astrophysics]
  {10.1051/0004-6361/201834311}, \href
  {https://ui.adsabs.harvard.edu/abs/2018A&A...620L..10C} {620, L10}

\bibitem[\protect\citeauthoryear{{Cantalloube}, {Dohlen}, {Milli}, {Brandner}
  \& {Vigan}}{{Cantalloube} et~al.}{2019}]{cantalloube19}
{Cantalloube} F.,  {Dohlen} K.,  {Milli} J.,  {Brandner} W.,   {Vigan} A.,
  2019, \mn@doi [The Messenger] {10.18727/0722-6691/5138}, \href
  {https://ui.adsabs.harvard.edu/abs/2019Msngr.176...25C} {176, 25}

\bibitem[\protect\citeauthoryear{Chew, Clare  \& Lane}{Chew
  et~al.}{2006}]{chew06}
Chew T.~Y.,  Clare R.~M.,   Lane R.~G.,  2006, \mn@doi [Optics Communications]
  {https://doi.org/10.1016/j.optcom.2006.07.011}, 268, 189

\bibitem[\protect\citeauthoryear{{Conan}}{{Conan}}{2003}]{conanr03}
{Conan} R.,  2003, Technical report, Fourier optics and distribution theory
  applied to pyramid wavefront sensors.
LAOG/ONERA

\bibitem[\protect\citeauthoryear{Conan \& Correia}{Conan \&
  Correia}{2014}]{conanr14}
Conan R.,  Correia C.,  2014, in SPIE 9148.

\bibitem[\protect\citeauthoryear{{Correia} \& {Teixeira}}{{Correia} \&
  {Teixeira}}{2014a}]{correia14}
{Correia} C.~M.,  {Teixeira} J.,  2014a, \mn@doi [Journal of the Optical
  Society of America A] {10.1364/JOSAA.31.002763}, \href
  {http://cdsads.u-strasbg.fr/abs/2014JOSAA..31.2763C} {31, 2763}

\bibitem[\protect\citeauthoryear{{Correia} \& {Teixeira}}{{Correia} \&
  {Teixeira}}{2014b}]{correia14a}
{Correia} C.~M.,  {Teixeira} J.,  2014b, \mn@doi [Journal of the Optical
  Society of America A] {10.1364/JOSAA.31.002763}, \href
  {http://cdsads.u-strasbg.fr/abs/2014JOSAA..31.2763C} {31, 2763}

\bibitem[\protect\citeauthoryear{{Correia}, {Conan}, {Kulcs{\'a}r}, {Raynaud},
  {Petit}  \& {Fusco}}{{Correia} et~al.}{2007}]{correia07}
{Correia} C.,  {Conan} J.~M.,  {Kulcs{\'a}r} C.,  {Raynaud} H.~F.,  {Petit} C.,
    {Fusco} T.,  2007, in {Bouvier} J.,  {Chalabaev} A.,   {Charbonnel} C.,
  eds, SF2A-2007: Proceedings of the Annual meeting of the French Society of
  Astronomy and Astrophysics. p.~25

\bibitem[\protect\citeauthoryear{Correia, Kulcsar, Conan  \& Raynaud}{Correia
  et~al.}{2008}]{Correia08}
Correia C.,  Kulcsar C.,  Conan J.-M.,   Raynaud H.-F.,  2008, in Hubin N.,
  Max C.~E.,   Wizinowich P.~L.,  eds,  Vol. 7015, Proc. of the SPIE. SPIE, p.
  701551, \mn@doi{10.1117/12.788455}, \url
  {http://link.aip.org/link/?PSI/7015/701551/1}

\bibitem[\protect\citeauthoryear{Correia, Bond, Sauvage, Fusco, Conan  \&
  Wizinowich}{Correia et~al.}{2017}]{correia17}
Correia C.~M.,  Bond C.~Z.,  Sauvage J.-F.,  Fusco T.,  Conan R.,   Wizinowich
  P.~L.,  2017, \mn@doi [J. Opt. Soc. Am. A] {10.1364/JOSAA.34.001877}, 34,
  1877

\bibitem[\protect\citeauthoryear{{Deo}, {Gendron}, {Rousset}, {Vidal}  \&
  {Buey}}{{Deo} et~al.}{2018}]{deo18}
{Deo} V.,  {Gendron} {\'E}.,  {Rousset} G.,  {Vidal} F.,   {Buey} T.,  2018, in
  Society of Photo-Optical Instrumentation Engineers (SPIE) Conference Series.
  p. 1070320, \mn@doi{10.1117/12.2311631}

\bibitem[\protect\citeauthoryear{Edl{\'{e}}n}{Edl{\'{e}}n}{1966}]{edlen66}
Edl{\'{e}}n B.,  1966, \mn@doi [Metrologia] {10.1088/0026-1394/2/2/002}, 2, 71

\bibitem[\protect\citeauthoryear{{Esposito}, {Pinna}, {Puglisi}, {Agapito},
  {Veran}  \& {Herriot}}{{Esposito} et~al.}{2015}]{esposito15}
{Esposito} S.,  {Pinna} E.,  {Puglisi} A.,  {Agapito} G.,  {Veran} J.~P.,
  {Herriot} G.,  2015, in Adaptive Optics for Extremely Large Telescopes IV
  (AO4ELT4). p.~E36

\bibitem[\protect\citeauthoryear{{Fauvarque}}{{Fauvarque}}{2017}]{fauvarque17a}
{Fauvarque} O.,  2017, PhD thesis, Aix-Marseille Univ

\bibitem[\protect\citeauthoryear{{Fauvarque}, {Neichel}, {Fusco}  \&
  {Sauvage}}{{Fauvarque} et~al.}{2015}]{fauvarque15}
{Fauvarque} O.,  {Neichel} B.,  {Fusco} T.,   {Sauvage} J.-F.,  2015, \mn@doi
  [Optics Letters] {10.1364/OL.40.003528}, \href
  {http://cdsads.u-strasbg.fr/abs/2015OptL...40.3528F} {40, 3528}

\bibitem[\protect\citeauthoryear{{Fauvarque}, {Neichel}, {Fusco}, {Sauvage}  \&
  {Girault}}{{Fauvarque} et~al.}{2017}]{fauvarque17}
{Fauvarque} O.,  {Neichel} B.,  {Fusco} T.,  {Sauvage} J.-F.,   {Girault} O.,
  2017, \mn@doi [Journal of Astronomical Telescopes, Instruments, and Systems]
  {10.1117/1.JATIS.3.1.019001}, \href
  {http://cdsads.u-strasbg.fr/abs/2017JATIS...3a9001F} {3, 019001}

\bibitem[\protect\citeauthoryear{{Fauvarque}, {Janin-Potiron}, {Correia},
  {Brule}, {Neichel}, {Chambouleyron}, {Sauvage}  \& {Fusco}}{{Fauvarque}
  et~al.}{2019}]{fauvarque19}
{Fauvarque} O.,  {Janin-Potiron} P.,  {Correia} C.,  {Brule} Y.,  {Neichel} B.,
   {Chambouleyron} V.,  {Sauvage} J.-F.,   {Fusco} T.,  2019, arXiv e-prints,
  \href {https://ui.adsabs.harvard.edu/\#abs/2019arXiv190205440F} {p.
  arXiv:1902.05440}

\bibitem[\protect\citeauthoryear{Feeney}{Feeney}{2001}]{feeney01}
Feeney O.~A.,  2001, Technical report, Theory and Laboratory Characterisation
  of Novel Wavefront Sensor for Adaptive Optics Systems.
National University of Ireland, Galway

\bibitem[\protect\citeauthoryear{{Fusco} et~al.,}{{Fusco}
  et~al.}{2006}]{fusco06}
{Fusco} T.,  et~al., 2006, \mn@doi [Optics Express] {10.1364/OE.14.007515},
  \href {https://ui.adsabs.harvard.edu/abs/2006OExpr..14.7515F} {14, 7515}

\bibitem[\protect\citeauthoryear{{Fusco} et~al.,}{{Fusco}
  et~al.}{2016}]{fusco16}
{Fusco} T.,  et~al., 2016, in Adaptive Optics Systems V. p. 99090U,
  \mn@doi{10.1117/12.2233319}

\bibitem[\protect\citeauthoryear{Guyon}{Guyon}{2005}]{guyon05}
Guyon O.,  2005, The Astrophysical Journal, 629, 592

\bibitem[\protect\citeauthoryear{{Guyon}}{{Guyon}}{2018}]{guyon18}
{Guyon} O.,  2018, \mn@doi [Annual Review of Astronomy and Astrophysics]
  {10.1146/annurev-astro-081817-052000}, \href
  {https://ui.adsabs.harvard.edu/abs/2018ARA&A..56..315G} {56, 315}

\bibitem[\protect\citeauthoryear{J.W.Hardy}{J.W.Hardy}{1998}]{hardy98}
J.W.Hardy 1998, Adaptive Optics for Astronomical Telescopes.
Oxford, New York

\bibitem[\protect\citeauthoryear{Jared R.~Males}{Jared
  R.~Males}{2018}]{males18}
Jared R.~Males O.~G.,  2018, \mn@doi [Journal of Astronomical Telescopes,
  Instruments, and Systems] {10.1117/1.JATIS.4.1.019001}, 4, 4

\bibitem[\protect\citeauthoryear{Korkiakoski, V\'{e}rinaud, Louarn  \&
  Conan}{Korkiakoski et~al.}{2007}]{korkiakoski07}
Korkiakoski V.,  V\'{e}rinaud C.,  Louarn M.~L.,   Conan R.,  2007, \mn@doi
  [Appl. Opt.] {10.1364/AO.46.006176}, 46, 6176

\bibitem[\protect\citeauthoryear{LeDue, Jolissaint, V\'{e}ran  \&
  Bradley}{LeDue et~al.}{2009}]{ledue09}
LeDue J.,  Jolissaint L.,  V\'{e}ran J.-P.,   Bradley C.,  2009, \mn@doi [Opt.
  Express] {10.1364/OE.17.007186}, 17, 7186

\bibitem[\protect\citeauthoryear{{Linfoot}}{{Linfoot}}{1948}]{linfoot48}
{Linfoot} E.~H.,  1948, \mn@doi [\mnras] {10.1093/mnras/108.6.428}, \href
  {http://adsabs.harvard.edu/abs/1948MNRAS.108..428L} {108, 428}

\bibitem[\protect\citeauthoryear{{Macintosh} et~al.,}{{Macintosh}
  et~al.}{2018}]{macintosh19}
{Macintosh} B.,  et~al., 2018, in \procspie. p. 107030K (\mn@eprint {arXiv}
  {1807.07146}), \mn@doi{10.1117/12.2314253}

\bibitem[\protect\citeauthoryear{Massioni, Gilles  \& Ellerbroek}{Massioni
  et~al.}{2015}]{massioni15}
Massioni P.,  Gilles L.,   Ellerbroek B.,  2015, \mn@doi [J. Opt. Soc. Am. A]
  {10.1364/JOSAA.32.002353}, 32, 2353

\bibitem[\protect\citeauthoryear{{Mawet} et~al.,}{{Mawet}
  et~al.}{2012}]{mawet12}
{Mawet} D.,  et~al., 2012, {Review of small-angle coronagraphic techniques in
  the wake of ground-based second-generation adaptive optics systems}.
p. 844204, \mn@doi{10.1117/12.927245}

\bibitem[\protect\citeauthoryear{{Mawet} et~al.,}{{Mawet}
  et~al.}{2014}]{mawet14}
{Mawet} D.,  et~al., 2014, \mn@doi [\apj] {10.1088/0004-637X/792/2/97}, \href
  {https://ui.adsabs.harvard.edu/abs/2014ApJ...792...97M} {792, 97}

\bibitem[\protect\citeauthoryear{{Mawet} et~al.,}{{Mawet}
  et~al.}{2016}]{mawet16}
{Mawet} D.,  et~al., 2016, {Keck Planet Imager and Characterizer: concept and
  phased implementation}.
p. 99090D, \mn@doi{10.1117/12.2233658}

\bibitem[\protect\citeauthoryear{{Mouillet} et~al.,}{{Mouillet}
  et~al.}{2018}]{mouillet18}
{Mouillet} D.,  et~al., 2018, in \procspie. p. 107031Q,
  \mn@doi{10.1117/12.2313277}

\bibitem[\protect\citeauthoryear{{Oppenheim} \& {Schafer}}{{Oppenheim} \&
  {Schafer}}{1999}]{oppenheim99}
{Oppenheim} A.~V.,  {Schafer} R.~W.,  1999, Discrete-time signal processing,
  2nd edn.
Prentice-Hall,Inc.

\bibitem[\protect\citeauthoryear{Owens}{Owens}{1967}]{owens67}
Owens J.~C.,  1967, \mn@doi [Appl. Opt.] {10.1364/AO.6.000051}, 6, 51

\bibitem[\protect\citeauthoryear{Poyneer \& V{\'e}ran}{Poyneer \&
  V{\'e}ran}{2005}]{Poyneer05}
Poyneer L.~A.,  V{\'e}ran J.-P.,  2005, Journal of the Optical Society of
  America A, 22, 1515

\bibitem[\protect\citeauthoryear{Quir{\'o}s-Pacheco, Correia  \&
  Esposito}{Quir{\'o}s-Pacheco et~al.}{2009}]{QuirosPacheco09}
Quir{\'o}s-Pacheco F.,  Correia C.,   Esposito S.,  2009, in Y.~Cl{\'e}net
  J.-M.~Conan T.~F.,  Rousset G.,  eds, 1st AO4ELT Conference - Adaptative
  Optics for Extremely Large Telescopes proceedings. No. 07005.
EDP Sciences

\bibitem[\protect\citeauthoryear{Ragazzoni}{Ragazzoni}{1996}]{ragazzoni96}
Ragazzoni R.,  1996, Journal of Modern Optics, 43, 289

\bibitem[\protect\citeauthoryear{Rigaut \& Gendron}{Rigaut \&
  Gendron}{1992}]{rigaut92}
Rigaut F.,  Gendron E.,  1992, Astronomy and Astrophysics, 261, 677

\bibitem[\protect\citeauthoryear{Roddier}{Roddier}{1981}]{roddier81}
Roddier F.,  1981, The effects of atmospherical turbulence in optical
  astronomy.
pp 281--376

\bibitem[\protect\citeauthoryear{{Shatokhina} \& {Ramlau}}{{Shatokhina} \&
  {Ramlau}}{2017}]{shatokhina17}
{Shatokhina} I.,  {Ramlau} R.,  2017, \ao, 56, 6381

\bibitem[\protect\citeauthoryear{Shatokhina, Obereder, Rosensteiner  \&
  Ramlau}{Shatokhina et~al.}{2013}]{shatokhina13}
Shatokhina I.,  Obereder A.,  Rosensteiner M.,   Ramlau R.,  2013, \mn@doi
  [Appl. Opt.] {10.1364/AO.52.002640}, 52, 2640

\bibitem[\protect\citeauthoryear{{Snik} et~al.,}{{Snik} et~al.}{2018}]{snik18}
{Snik} F.,  et~al., 2018, in \procspie. p. 107062L (\mn@eprint {arXiv}
  {1807.07100}), \mn@doi{10.1117/12.2313957}

\bibitem[\protect\citeauthoryear{{Thomas}, {Fusco}, {Tokovinin}, {Nicolle},
  {Michau}  \& {Rousset}}{{Thomas} et~al.}{2006}]{thomas06}
{Thomas} S.,  {Fusco} T.,  {Tokovinin} A.,  {Nicolle} M.,  {Michau} V.,
  {Rousset} G.,  2006, \mn@doi [\mnras] {10.1111/j.1365-2966.2006.10661.x},
  371, 323

\bibitem[\protect\citeauthoryear{V{\'e}rinaud}{V{\'e}rinaud}{2004}]{Verinaud04}
V{\'e}rinaud C.,  2004, Optics Communications, 233, 27

\bibitem[\protect\citeauthoryear{V{\'e}rinaud, {Le Louarn}, Korkiakoski  \&
  Carbillet}{V{\'e}rinaud et~al.}{2005}]{verinaud05}
V{\'e}rinaud C.,  {Le Louarn} M.,  Korkiakoski V.,   Carbillet M.,  2005,
  Monthly Notices of the Royal Astronomical Society, 357, L26

\bibitem[\protect\citeauthoryear{{Vigan} et~al.,}{{Vigan}
  et~al.}{2019}]{vigan19}
{Vigan} A.,  et~al., 2019, \mn@doi [\aap] {10.1051/0004-6361/201935889}, \href
  {https://ui.adsabs.harvard.edu/abs/2019A&A...629A..11V} {629, A11}

\bibitem[\protect\citeauthoryear{Wang, Bai, Ning, Huang  \& Wang}{Wang
  et~al.}{2010}]{wang10}
Wang J.,  Bai F.,  Ning Y.,  Huang L.,   Wang S.,  2010, \mn@doi [Opt. Express]
  {10.1364/OE.18.027534}, 18, 27534

\bibitem[\protect\citeauthoryear{{Xuan} et~al.,}{{Xuan} et~al.}{2018}]{xuan18}
{Xuan} W.~J.,  et~al., 2018, \mn@doi [\aj] {10.3847/1538-3881/aadae6}, \href
  {https://ui.adsabs.harvard.edu/abs/2018AJ....156..156X} {156, 156}

\makeatother
\end{thebibliography}




\appendix
\section{Noise propagation expressed on an orthonormal basis of modes}
Results in this section follow closely the approach of
\cite{rigaut92}. 
The propagated noise covariance matrix on a predefined basis set of
modes is defined as
\begin{equation}
\CovMat_m = \average{\mathbf{m} \mathbf{m}^\T}
\end{equation}
with the reconstructed modal coefficient vector $\mathbf{m}$ 
 \begin{equation}
\mathbf{m} = \D^\dag \etavec
\end{equation}
where $\etavec$ is a vector WFS measurement signals,  $\D^\dag = (\D \D^\T)^{-1}\D^\T$ and $\D$ is the
system interaction matrix $\svec = \D \phivec$ containing the responses of the WFS to each mode
in $\mathbf{m}_i$ of $\phivec$. For the PWFS this is the covariance function that
needs be evaluated, i.e.
\begin{align}
\CovMat_m & = \D^\dag\average{\etavec \etavec^\T}\D^\dag
\end{align}
and $\average{\etavec \etavec^\T}$
provided in Eq. (\ref{eq:sigmaPhNoiseMeasurementPWFS_final})  and
Eq. (\ref{eq:sigmaRONMeasurementPWFS_final}). 

For linear WFSs a simplification applies. If we consider now a measurement of pure noise which is assumed of
constant variance 
$\sigma^2_\eta$ across the pupil then the noise propagated is 
\begin{align}
\CovMat_m & = \D^\dag\average{\svec \svec^\T}\D^\dag\\
 & = (\D \D^\T)^{-1} \sigma^2_\eta
\end{align}
provided that $\average{\svec \svec^\T} = \I \sigma^2_\eta$. Using SVD
decomposition
\begin{equation}
[\mathcal{U} \mathcal{S} \mathcal{V}^\T] = svd(\D)
\end{equation}
where $\mathcal{S} = diag(\beta_{ii})$ is a diagonal matrix with the
singular values in it.

One can likewise write
\begin{equation}
 (\D \D^\T)^{-1} = \left( \mathcal{V} \mathcal{S}^\T \mathcal{U}^\T \mathcal{U} \mathcal{S} \mathcal{V}^\T\right)^{-1}
\end{equation}
Using the equalities $ \mathcal{U}^\T= \mathcal{U}^{-1}$ and $
\mathcal{V}^\T= \mathcal{V}^{-1}$
one gets
\begin{equation}
 (\D \D^\T)^{-1} =  \mathcal{V} \mathcal{S}^{-\T}\mathcal{S}^{-1} \mathcal{V}^\T
\end{equation}
from which the $k^{th}$ element of the diagonal writes
\begin{align}
 \sigma^2_k & =  \sum_i \mathcal{V}_{k,i}
              \frac{1}{\beta_{ii}^2}\mathcal{V}_{ik}^\T\\
 & = \sum_i \mathcal{V}_{ki}
              \frac{1}{\beta_{ii}^2}\mathcal{V}_{ki}\\
& = \sum_i \frac{\mathcal{V}_{ki}^2}{\beta_{ii}^2}
\end{align}
which is a more straightforward demonstration but otherwise equivalent
to that of \cite{feeney01}. 

We caution the reader however that the PWFS noise propagation coefficients cannot be quoted in units of angle-on-sky since its measurements, unlike the SH-WFS, are not straight wave-front gradients (or slopes). 

\section{Variations around WFS and controller choices}
\begin{figure}
	\begin{center}
            \includegraphics[width=0.5\textwidth]{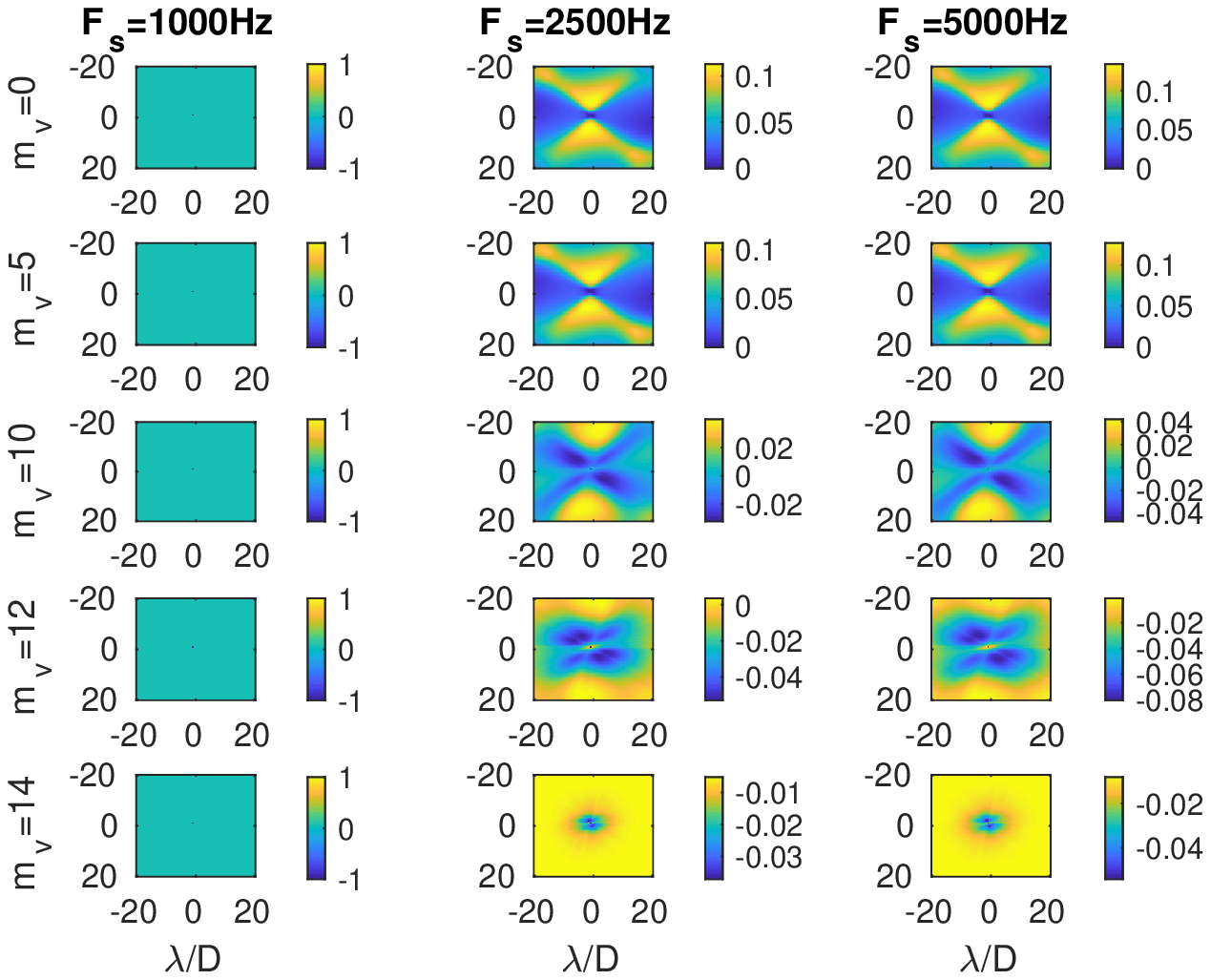}
	\end{center}
	\caption[]
	{\label{fig:sphere2p0ContrastRatio_SH_DKF_vs_Ts}
Contrast ratio map as a function of the frame-rate and stellar
magnitude for a SH with a DKF controller.}
\end{figure}

\begin{figure}
	\begin{center}
            \includegraphics[width=0.5\textwidth]{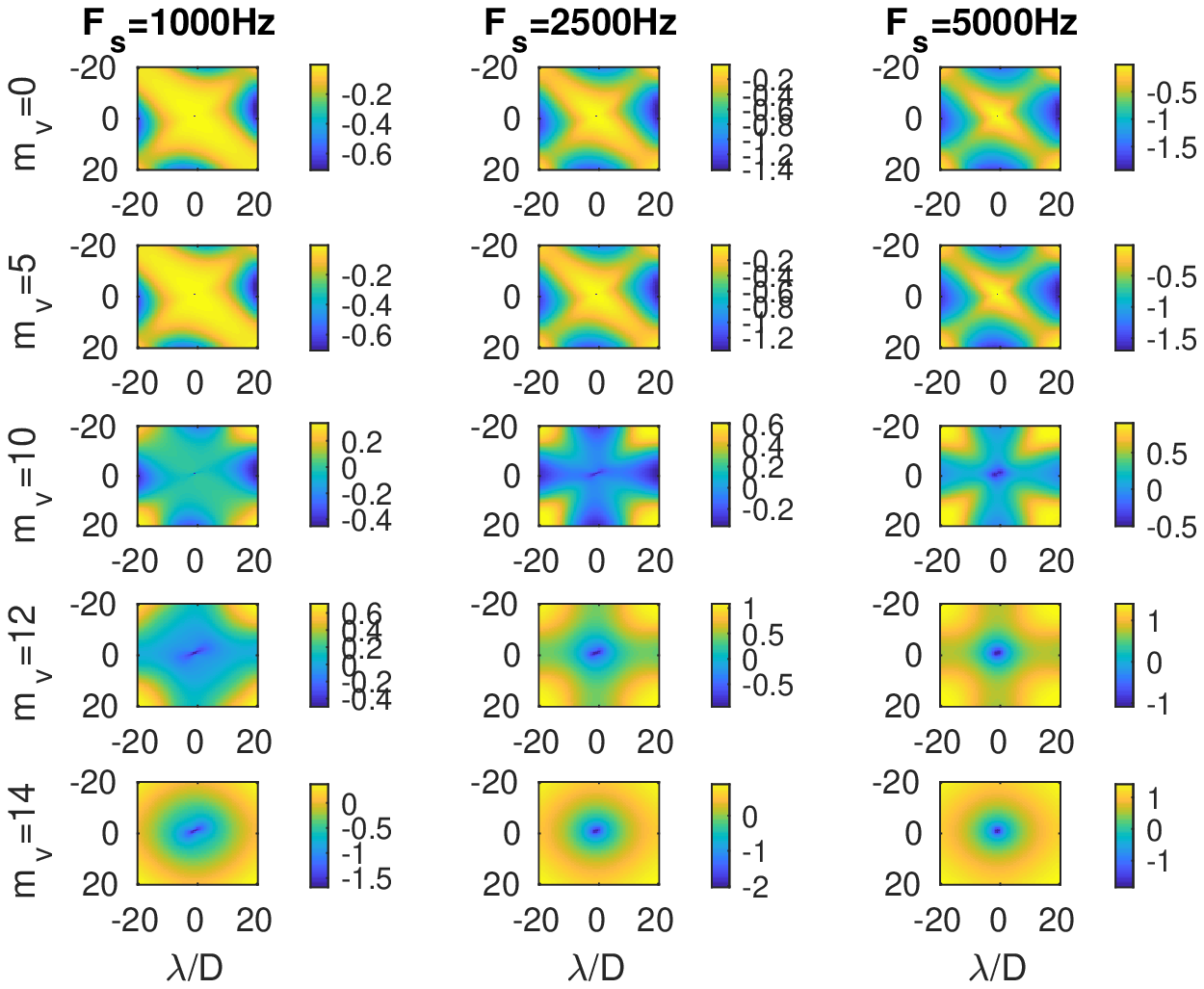}
	\end{center}
	\caption[]
	{\label{fig:sphere2p0ContrastRatio_PWFS_LS_vs_SH_LS} Contrast ratio map as a function of the frame-rate and stellar
magnitude for a PWFS versus a SH-based system both with a LS
reconstructor and integral controller.}
\end{figure}

\begin{figure}
	\begin{center}
            \includegraphics[width=0.5\textwidth]{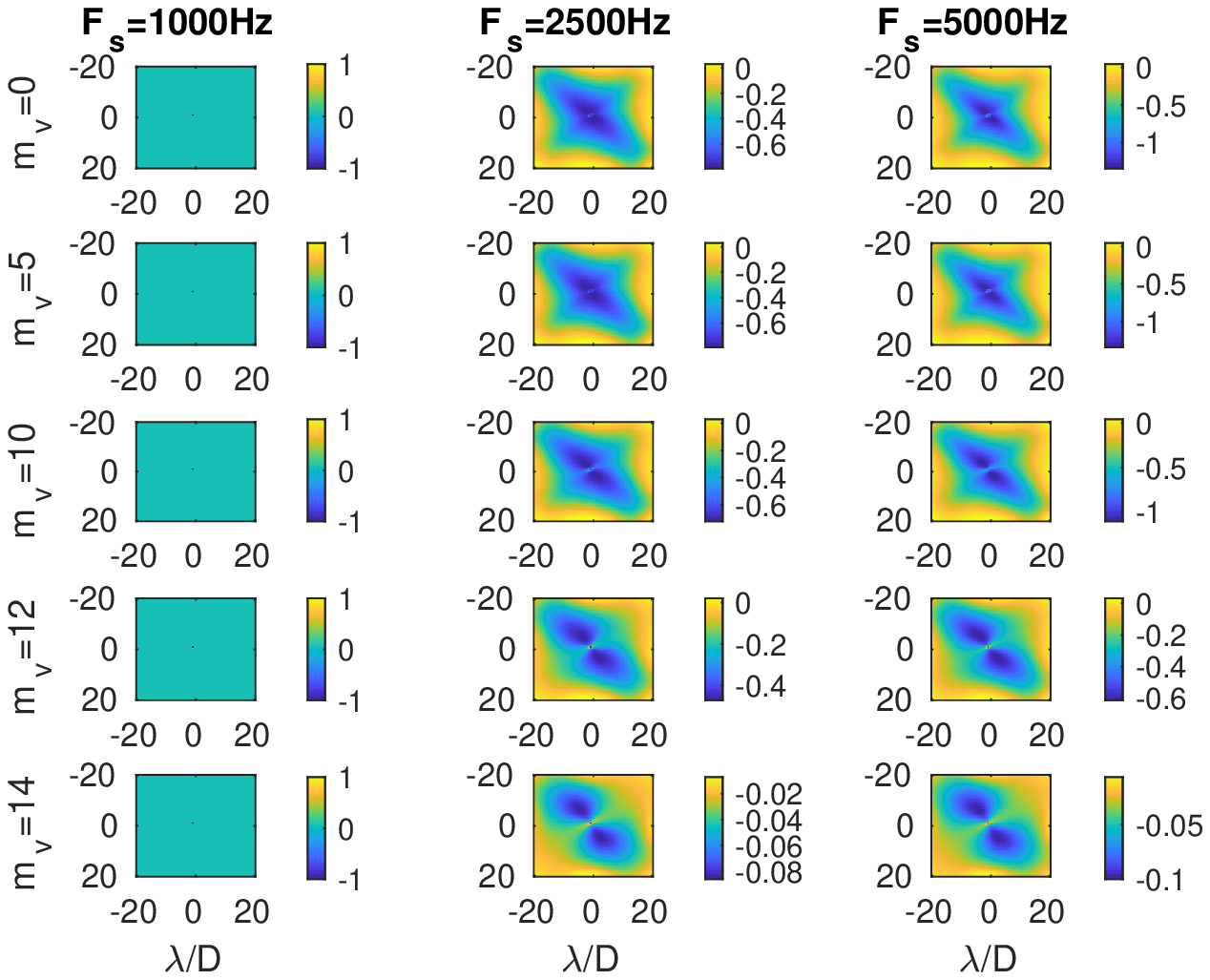}
	\end{center}
	\caption[]
	{\label{fig:sphere2p0ContrastRatio_SH_LS_vs_Ts}
Contrast ratio map as a function of the frame-rate and stellar
magnitude for a SH with a DKF controller.}
\end{figure}

Figures \ref{fig:sphere2p0ContrastRatio_SH_DKF_vs_Ts},
\ref{fig:sphere2p0ContrastRatio_PWFS_LS_vs_SH_LS} and \ref{fig:sphere2p0ContrastRatio_SH_LS_vs_Ts} depict the contrast ratios
obtained as a function of the stellar magnitude and frame-rate for the
different choices of WFS and controller (for a constant pure loop
delay of 3 ms).

The remainder of the figures show different combinations of
controllers and WFS. 


Figure \ref{fig:sphere2p0ContrastRatio_SH_DKF_vs_Ts} shows the
contrast improvements by increasing the frame rate for a SH-based
system with the DKF controller. Gains are observed in a butterfly
shaped region for brighter stars and it vanished as noise becomes the
dominant factor for fainter stars. 

If we now look into the contrast gains of just replacing the SH by a
PWFS, but keeping the controller we find results in
Fig. \ref{fig:sphere2p0ContrastRatio_PWFS_LS_vs_SH_LS}. It seems
that only a poor improvement is achieved regardless of the magnitude
and frame-rate chosen.

Finally, Fig. \ref{fig:sphere2p0ContrastRatio_SH_LS_vs_Ts} shows the
improvement of increasing the frame-rate with the SH and integral
controller. Compared to Fig. \ref{fig:sphere2p0ContrastRatio_SH_DKF_vs_SH_LS}  the contrast improvements
are not nearly as spectacular since the controller adds no predictive
knowledge to the wave-front estimation in order to further improve
contrast as small separations. 

\section*{Acknowledgements}
All the simulations and analysis were done with the object- oriented Matlab AO simulator (OOMAO) [30]. The class spatialFrequencyAdaptiveOptics implementing the analytics developed in this paper as well as the results herein is packed with the end-to-end library freely available from \href{https://github.com/cmcorreia/oomao}{https://github.com/cmcorreia/oomao}.

The research leading to these results received the support of the A*MIDEX project (no. ANR-11-IDEX-0001- 02) funded by the ”Investissements d'Avenir” French Government program, managed by the French National Research Agency (ANR).


\bsp	
\label{lastpage}
\end{document}